\documentclass[%
 reprint,
 amsmath,amssymb,
 aps,
nofootinbib,
superscriptaddress,
showkeys
]{revtex4-1}

\usepackage{graphicx}
\usepackage{dcolumn}
\usepackage{bm}

\usepackage{mathtools}
\usepackage[usenames, dvipsnames]{xcolor}

\newcommand{\eref}[1]{(\ref{#1})}
\usepackage{tikz}
\usepackage{nicefrac}
\usepackage[caption=false]{subfig}
\usepackage{mathptmx}

\usepackage[]{hyperref}
\hypersetup{
    colorlinks,%
    citecolor=blue,%
    filecolor=black,%
    linkcolor=blue,%
    urlcolor=black
}

\usepackage{cleveref}
\usepackage{tabularx}

\begin{document}
\title{Mechanism of Re precipitation in irradiated W-Re alloys from kinetic Monte Carlo simulations}

\author{Chen-Hsi Huang}
\affiliation{Department of Materials Science and Engineering, University of California, Los Angeles, CA 90095, USA}
\author{Leili Gharaee}
\affiliation{Department of Physics, Chalmers University of Technology, S-412 96 Gothenburg, Sweden}
\author{Yue Zhao}
\affiliation{Department of Materials Science and Engineering, University of California, Los Angeles, CA 90095, USA}
\author{Paul Erhart}
\affiliation{Department of Physics, Chalmers University of Technology, S-412 96 Gothenburg, Sweden}
\author{Jaime Marian}
\affiliation{Department of Materials Science and Engineering, University of California, Los Angeles, CA 90095, USA}
\affiliation{Department of Mechanical and Aerospace Engineering, University of California, Los Angeles, CA 90095, USA}



\begin{abstract}
High-temperature, high-dose, neutron irradiation of W results in the formation of Re-rich clusters at concentrations one order of magnitude lower than the thermodynamic solubility limit. These clusters may eventually transform into brittle W-Re intermetallic phases, which can lead to high levels of hardening and thermal conductivity losses. Standard theories of radiation enhanced diffusion and precipitation cannot explain the formation of these precipitates and so understanding the mechanism by which nonequilibrium clusters form under irradiation is crucial to predict materials degradation and devise mitigation strategies. 
Here we carry out a thermodynamic study of W-Re alloys and conduct kinetic Monte Carlo simulations of Re cluster formation in irradiated W-2Re alloys using a generalized Hamiltonian for crystals containing point defects parameterized entirely with electronic structure calculations. Our model incorporates recently-gained mechanistic  information of mixed-interstitial solute transport, which is seen to control cluster nucleation and growth by forming quasi-spherical nuclei after an average incubation time of 20 s at 1800 K. These nuclei are seen to grow by attracting more mixed interstitials bringing solute atoms, which in turns attracts vacancies leading to recombination and solute agglomeration. The clusters grow to a maximum size of approximately 4-nm radius, and are not fully dense with Re, containing 50\% or less near the center.  Our simulations are in reasonable agreement with recent atom probe examinations of ion irradiated W-2Re systems at 773 K.
\end{abstract}

\keywords{W-Re alloys; Solute precipitation; Neutron irradiation; kinetic Monte Carlo}

\maketitle

\section{Introduction}\label{sec:intro}
Tungsten is the prime candidate material in magnetic fusion energy devices due to its high strength and excellent high temperature properties \cite{ZinGho00,RieDudGon13,FitNgu08,BecDom09}. Upon fast neutron irradiation in the 600-1000$^\circ$C temperature range, W transmutes into Re by the way of beta decay reactions at a rate that depends on the neutron spectrum and the position in the reactor. For the DEMO (demonstration fusion power plant) reactor concept, calculations show that the transmutation rate is 2000 and 7000 atomic parts per million (appm) per displacements per atom (dpa) in the divertor and the equatorial plane of the first wall, respectively (where damage, in each case, accumulates at rates of 3.4 and 4.4 dpa/year) \cite{GilSub11,GilDudZhe12}.
The irradiated microstructure initially evolves by accumulating a high density of prismatic dislocation loops and vacancy clusters, approximately up to 0.15 dpa \cite{he2006microstructural,TanHasHe09,HasTanNog11,Hu2016235}. Subsequently, a void lattice emerges and fully develops at fluences of around 1 dpa. After a critical dose that ranges between 5 dpa for fast ($>$1 MeV) neutron irradiation \cite{HasTanNog11} and 2.2 dpa in modified target rabbits in the HFIR \cite{HuKoyKat15,Hu2016235}, W and W-Re alloys develop a high density of nanometric precipitates with acicular shape at Re concentrations well below the solubility limit \cite{HasTanNog11,Hu2016235}. The structure of these precipitates is consistent with $\sigma$ (W$_7$Re$_6$) and $\chi$ (WRe$_3$) intermetallic phases, which under equilibrium conditions only occur at temperatures and Re concentrations substantially higher than those found in neutron irradiation studies \cite{cottrell2004sigma}. A principal signature of the formation of these intermetallic structures in body-centered cubic (bcc) W is the sharp increase in hardness and embrittlement \cite{TanHasHe09,HasTanNog11,Hu2016235}. Qualitatively similar observations have been recently made in W-2Re and W-1Re-1Os alloys subjected to heavy ion irradiation \cite{XuBecArm15,2017ACTAXu}, clearly establishing a link between primary damage production and Re precipitation.

Precipitation of nonequlibrium phases in irradiated materials is commonplace. The standard theory of irradiation damage includes radiation enhanced diffusion (RED) and radiation induced precipitation (RIP) as mechanisms that can drive the system out of equilibrium due to the onset of point defect cluster fluxes towards defect sinks \cite{was2007fundamentals,dienes1958radiation,Cauvin197967}. Within this picture, Re precipitation in W or W-Re alloys under irradiation would then, in principle, be unsurprising were it not for the fact that Re clustering is seen to occur at concentrations still below the solubility limit even after RED has taken place. In spite of this, recent work using energy models based on the cluster expansion formalism for the W-Re system, and fitted to density functional theory (DFT) calculations, have revealed a direct relationship between excess vacancy concentrations and the formation of Re solute-rich clusters \cite{2016arXiv160403746W}. These calculations are substantiated by recent neutron irradiation experiments of pure W at 900$^{\circ}$C up to 1.6 dpa in the HFR in Petten \cite{Klimenkov2016}. Post-irradiation examination of the irradiated specimens reveals the formation of a fine distribution of voids with average 5 nm size surrounded by Re-rich clouds. However, the relative concentration of Re around the voids is still on the order of 12-18\% (from a nominal overall concentration of 1.4\% from transmutation), well below the precipitation limit of Re in W at 900$^{\circ}$C. However, in the ion beam irradiation experiments of W-2Re alloys by Xu {\it et al.}~at 300 and 500$^{\circ}$C, Re-rich clusters with bcc structure are seen to form with concentrations between 12 and 30\% Re with no indication of vacancies forming part of the clusters \cite{XuBecArm15,2017ACTAXu}. Another piece of evidence against a strong association between vacancies and Re atoms comes from irradiation tests of W-Re alloys performed at EBR-II in the 1970s and 80s \cite{Matolich1974837,sikka1974identification,WilWifBen83,herschitz1984atomic,herschitz1985radiation}. In these studies, the presence of Re was seen to suppress swelling, which would seem to suggest a decoupling between vacancy clusters and Re atoms.
Clearly, equilibrium thermodynamics involving vacancies alone may not suffice to explain the precipitation tendencies in irradiated W-Re alloys. 

All this is suggestive of alternative solute transport mechanisms that may be unique to W-Re systems. 
Indeed, several recent studies using electronic structure calculations have independently reported a peculiar association between self-interstitial atoms (SIA) and Re solutes that results in very high solute transport efficacy \cite{2014MSMSESuzudo,2015JNMSuzudo,2016JAPGharaee}. This mechanism consists of a series of mixed dumbbell rotations and translations such that the mixed nature of the dumbbell is preserved and solutes can be transported over long distances without the need for vacancy exchanges. Furthermore, this mechanism effectively transforms one-dimensional SIA diffusion into a 3D mixed-dumbbell transport process at activation energies considerably lower than that of vacancy diffusion. The objective of this paper is to study the kinetics of Re precipitation in irradiated W accounting for both vacancy and mixed-interstitial solute transport. To this effect, we develop a lattice kinetic Monte Carlo (kMC) model of alloy evolution parameterized solely using first principles calculations. We start in Section \ref{sec:model} by describing the essential elements of our kinetic model as well as the parameterization effort based on DFT calculations. In Section \ref{sec:res} we provide our main results, including semi-grand canonical Monte Carlo calculations of ternary W-Re-vacancy and W-Re-SIA systems, and kMC simulations and analysis of the Re-precipitate nucleation and growth. We finish with a discussion of the results and the conclusions in Section \ref{sec:disc}.

\section{Theory and methods}\label{sec:model}
\subsection{Energy model }

The energy model employed throughout this work is a cluster expansion Hamiltonian based on pair interactions truncated to the  $2^{\rm nd}$-nearest neighbor (2nn) shell:
\begin{equation} \label{eq_hamiltonian}
	{\cal H}= \sum_i\sum_{\alpha, \beta} n_{\alpha \text{-} \beta}^{(i)} \epsilon_{\alpha \text{-} \beta}^{(i)}  
\end{equation}
where $(i)$ specifies the type of nearest-neighbor interaction (first or second), $\alpha$ and $\beta$ refer to a pair of lattice sites, separated by a distance specified by the index $i$, $n_{\alpha \text{-} \beta}$ denotes the number of occurrences (bonds) of each $\alpha \text{-} \beta$ pair, and $\epsilon_{\alpha \text{-} \beta}$ is bond energy. 
In a previous work, we have shown how this Hamiltonian can be reduced to a generalized Ising Hamiltonian involving solvent and solute atoms (A and B), vacancies (V), and pure and mixed interstitials (AA, BB, and AB)  \cite{2016JPCMHuang}. The Hamiltonian is then expressed as a sum of polynomial terms of various degrees involving spin variables $\sigma_\alpha$ and $\sigma_\beta$ in the manner of the Ising model:
\begin{equation} \label{eq2_hamiltonian}
	{\cal H}= \sum_{n,m}\sum_{\alpha, \beta} \sigma^n_\alpha \sigma^m_\beta
\end{equation}
One of the advantages of using this notation is that the values assigned to the spin variables conserve the number of atoms $N$ of the system. We refer the reader to ref.\ \cite{2016JPCMHuang} for more details about this notation. In this paper we focus on the parameterization exercise for irradiated W-Re alloys\footnote{With A: W atoms; B: Re atoms; V: vacancies, AA: W-W dumbbell (or self-interstitial atom); BB: Re-Re dumbbell; AB: mixed W-Re dumbbell.}.

\subsection{Semi-Grand Canonical Monte Carlo for AB systems} \label{sec_mcAB}
The thermodynamic phase diagram of the W-Re system can be studied using semi-grand canonical Monte Carlo (SGMC) calculations as a function of temperature and solute concentration \cite{1981ACTABinder, 1993RRBDunweg, 1999ACTAPareige, 2004PRBTavazza, 2005PRBCannavacciuolo, 2010INTMETALBiborski}. We seek to minimize the thermodynamic potential of the semi-grand canonical ensemble, characterized by a constant temperature $T$, a constant number of particles $N$, and a constant chemical potential $\mu$. In each SGMC step, an A atom is randomly flipped into a B atom and the new state is accepted with probability:
\begin{equation} \label{eq_GCensemble}
	p_{ij}= \exp \left(-\frac{\Delta {\cal H}_{ij} - N_{\rm B}\Delta \mu }{k_BT} \right)
\end{equation}
where $\Delta {\cal H}_{ij}$ is the energy difference between the initial and final states, $i$ and $j$, $N_{\rm B}=NX$ is the number of solute atoms ($X$: solute concentration), $\Delta \mu$ is the change in chemical potential per atom after the transition, and $k_B$ is Boltzmann's constant. In this work, each transition is defined by changing the chemical nature of one atom chosen at random. In terms of the change in spin variable (in the notation of the generalized Ising Hamiltonian,  cf.\ eq.\ \eqref{eq2_hamiltonian}), this always results in a change of $\delta\sigma=-2$, such that eq.\ \eref{eq_GCensemble} can be simplified to:
\begin{equation} \label{eq_SGCprob}
	p_{ij}= \exp \left(-\frac{\Delta {\cal H}_{ij} +2\Delta\mu }{k_BT} \right)
\end{equation}

In the calculations, the chemical potential difference $\Delta\mu$ and the temperature $T$ are input variables, while the solute composition $X$ and the equilibrium configurations are obtained when convergence is reached.

\subsection{Metropolis Monte Carlo calculations of ABV system configurations} \label{sec_mcABV}
During irradiation, the introduction of large amounts of defects has the potential to impact the thermodynamics of the system. It is therefore of interest to calculate phase diagrams with fixed defect concentrations using equilibrium (Metropolis) Monte Carlo.  Defect concentrations are not thermodynamically equilibrated under irradiation --the number of vacancies or interstitials is not controlled by the chemical potential--, and so the AB system must be considered in conjunction with a fixed defect concentration. Take the case of vacancies for example, to properly obtain  converged nonequilibrium configurations of ABV systems, we employ a \textit{flip} and {\it swap} approach: (i) initially a system consisting of A atoms and a random distribution of vacancies is considered; (ii) a lattice point is selected at random; (iii) if that lattice point corresponds to an atom, a SGMC step is carried out, resulting in a change in the relative concentrations of A and B; if it, on the contrary, corresponds to a vacant site, then a canonical Monte Carlo step is carried out, leaving $X$ unchanged, and the vacancy is swapped with a randomly-selected atom. This trial swap is then accepted according to the Boltzmann distribution:
\begin{equation}
	p_{ij}= \exp \left(-\frac{\Delta {\cal H}_{ij}}{k_BT} \right)
\end{equation}
In this fashion, equilibrated AB alloys containing a fixed vacancy concentration are obtained, from which one can determine the changes relative to the thermodynamic equilibrium configurations. Although interstitials are much higher in energy than vacancies (so that only very small concentrations need be explored), the procedure for the ABI system is identical to that of the ABV system.

\subsection{Kinetic Monte Carlo simulations of ABVI systems} \label{sec_mcABVI}

The kinetic evolution of W-Re alloys under irradiation is studied using standard lattice kMC. 
The system is evolved by events involving atomic jumps and time is advanced according to the residence-time algorithm \cite{1966PPSYoung}.  Jump rates are calculated as:
\begin{equation} \label{eq_rate}
	r_{ij}= \nu \exp \left(-\frac{\Delta E_{ij}}{k_BT} \right)
\end{equation} 
where $\nu$ is an attempt frequency and $\Delta E_{ij}$ is the activation energy to jump from state $i$ to state $j$. 

\subsubsection{Vacancy migration model}
Several models have been proposed to describe the activation energy based on different interpretations of the atomic migration process (see, e.g.~\cite{2010JoNMSoisson} and \cite{2016JPCMHuang} for recent reviews). In this work, the activation energy of vacancy jump is modeled by the saddle-point energy model (or cut-bond model) \cite{1996AMSoisson, 2007PRBSoisson, 2012PRBMartinez, 2016ActaSenninger}, according to which $E_{ij}$ is given by the energy difference of the configuration when the jumping atom is at saddle point and the initial configuration:
\begin{equation} \label{eq_sp}
	\Delta E_{ij}= \sum_{p} \epsilon^{sp}_{\alpha\text{-}p} - \sum_{q} \epsilon_{\alpha\text{-}q}^{(i)} - \sum_{r\neq \alpha} \epsilon_{\text{V-}r}^{(i)} + \sum \Delta E_{ij}^{\rm non\text{-}broken}
\end{equation}
where $\alpha$ is the jumping atom, V is the vacancy, and $\epsilon^{sp}$ are the bond energies between the atom at the saddle point and the neighboring atoms. The first term on the r.h.s.\ of eq.\ \eref{eq_sp} reflects the energy of the jumping atom at the saddle point. In this work, we consider interactions up to 2nn distances for this term\footnote{In the saddle-point configuration for vacancy migration, there are six 1nn bonds and six 2nn bonds, compared with eight and six for a lattice point configuration.}. The second and third terms on the r.h.s.\ of eq.\ \eref{eq_sp} are the energies of the jumping atom and the vacancy at the initial state $i$. Finally, the fourth term gives the energy difference between state $i$ and $j$ for the non-broken bonds due to local solute concentration changes. The dependence of bond coefficients on local solute concentration will be discussed in Section \ref{sec_par}.

\subsubsection{Interstitial defect migration model} 
Here we consider self-interstitial atoms of the AA type, and mixed-interstitials AB. Due to their rarity, BB interstitials are omitted in our calculations. In bcc metals, AA SIAs are known to migrate athermally in one dimension along $\langle 111\rangle$ directions, with sporadic rotations to other $\langle 111\rangle$ orientations. These processes, however, are treated separately. 
In contrast to vacancy migration, activation energies of interstitial jumps are calculated using the direct final-initial system energy model \cite{2007NIMPRDjurabekova, 2008JoNMVincent, 2008JoNMVincenta, 2011PRBReina}: 
\begin{equation}
	\Delta E_{ij}=
	\begin{cases}
		E_m+\Delta {\cal H}_{ij}, & \text{if}~ \Delta {\cal H}_{ij} > 0 \\
		E_m,~\quad\quad\quad  & \text{if}~ \Delta {\cal H}_{ij} < 0
	\end{cases}
\end{equation}
In addition, we include a bias due to the well-known phenomenon of \emph{correlation}, by which a forward jump is slightly more likely to occur than a backward jump. This is reflected in a correlation factor $f$ computed as the ratio of forward to backward jumps \cite{2013JNMZhou}, which in our simulations is temperature dependent.
Rotations between $\langle 111\rangle$ directions are simply characterized by an activation energy equal to a rotation energy $E_r$.

For their part, as pointed out in Section \ref{sec:intro}, recent DFT studies have revealed a new migration mechanism for mixed dumbbells in W alloys. This mechanism involves an non-dissociative sequence of rotations and translations such that the solute atom is always part of the mixed dumbbell (in contrast with the \emph{intersticialcy} or 'knock-on' mechanism commonly displayed by SIAs) \cite{2014MSMSESuzudo, 2015JNMSuzudo, 2015JNMGharaee, 2016JAPGharaee}. This  effectively makes AB interstitials move in three dimensions with 2nn jumps along $\langle 100\rangle$ directions. Calculations for the W-Re system have shown that the migration energy in this case is very low, on the order of one tenth of an eV. As we shall see, this has extraordinary implications on the kinetic evolution of irradiated W-Re alloys.

\subsubsection{Spontaneous events: recombination and absorption}\label{sec:spont}

Any recombination event occurs spontaneously (no sampling involved) when the distance between an interstitial defect and a vacancy is within the $3^{\rm rd}$ nearest neighbor shell. Another reaction considered to be instantaneous is the transition of a SIA into an AB dumbbell when it encounters a solute atom: AA+B$\rightarrow$AB+A. This is because the binding energy between a SIA and a Re solute atom has been calculated to be $-0.8$ eV (negative binding energies represent attraction). The distance for this transformation is set to be equal to the 1nn separation.

Defect absorption represents another type of spontaneous event. Absorption can occur at sinks, such as a plane located in a stationary position within the simulation box \cite{2006JoNMSoisson}, or a free surface \cite{2016JPCMHuang}. Sinks can potentially act also as defect emitters, as in the case of grain boundaries, dislocations, and free surfaces in real microstructures. Details about the implementation of these processes can be found in ref.\ \cite{2016JPCMHuang}.

\subsubsection{Frenkel pair generation}

In this work, defects are generated as Frenkel pairs at a prescribed rate set by the damage rate. To insert a defect pair, two atomic sites are chosen at random, one is replaced by a vacancy and the other with an interstitial formed by an A atom and the lattice atom.

\subsection{Parameters} \label{sec_par}

There are five distinct atomic species used in this work: W atoms (A), Re atoms (B), vacancies (V), SIAs (AA), and mixed-interstitials (AB). As mentioned above, our energy model consists of pairwise interactions up to the 2nn shell. After discounting interstitial-vacancy bonds, this amounts to 26 different types of bonds (13 for each nearest neighbor shell), all of which must be obtained using first-principles calculations. 
Moreover, as discussed by Martinez {\it et al.}~and Senninger {\it et al.}~ \cite{2012PRBMartinez,2016ActaSenninger}, several of these bond energies are sensitive to the local solute concentration and must be computed on the fly in each Monte Carlo step. Following Warczok {\it et al.} \cite{2012CMSWarczok}, we reduce the number of unknowns from 26 to 13 by partitioning bond energies according the following relation:
\begin{equation} \label{eq_e2e1}
	\epsilon^{(2)}= \epsilon^{(1)} \left(\frac{r_{\rm 2nn}}{r_{\rm 1nn}}\right)^{-6}
\end{equation}
which is used unless both bond energies can be explicitly calculated. For the bcc lattice, this results in $\epsilon_{\alpha\text{-}\beta}^{(1)}/\epsilon_{\alpha\text{-}\beta}^{(2)}= 0.421875$ for regular bond coefficients, and $\epsilon_{\alpha\text{-}\beta}^{sp(1)}/\epsilon_{\alpha\text{-}\beta}^{sp(2)}= 0.194052$ for saddle-point bond coefficients.

The local solute concentration is always computed up to the 2nn shell. Next we describe the parameterization procedure for each set of bond energies.

\subsubsection{W-Re parameters}
The W-Re bond coefficients are $\epsilon_{\text{A-A}}$, $\epsilon_{\text{B-B}}$, and $\epsilon_{\text{A-B}}$. They determine the thermodynamic equilibrium phase diagram of the alloy. The bond coefficients of $\epsilon_{\text{A-A}}$ and $\epsilon_{\text{B-B}}$ are obtained from the cohesive energy:
\begin{equation} \label{eq_coh}
    \begin{aligned}	E_{coh}^\text{A}= -\frac{z_1}{2} \epsilon_{\text{A-A}}^{(1)} - \frac{z_2}{2} \epsilon_{\text{A-A}}^{(2)} \\
	E_{coh}^\text{B}= -\frac{z_1}{2} \epsilon_{\text{B-B}}^{(1)} - \frac{z_2}{2} \epsilon_{\text{B-B}}^{(2)}
   \end{aligned}
\end{equation}
where $z_1$ and $z_2$ are coordination numbers for the 1nn and 2nn shells, respectively. The cohesive energies calculated using DFT are given in Table \ref{dftt}.
\begin{table}[h]
	\caption{Energetics of W-Re systems calculated with DFT.}
	\label{dftt}
	\centering
	\begin{tabular}{| c | c | c | c |}
	\hline
Quantity & Value & Source \\
	\hline
	\hline
$E_{coh}^\text{A}$ & $8.3276$ & This work	\\ \hline
$E_{coh}^\text{B}$ & $7.4070$ & This work	\\ \hline
$\Delta H^{\rm mix}$ & $-0.1571-0.2311 X$ & Ref.\ \cite{2016JAPGharaee} 	\\ \hline
$E_f^\text{V}$ & $3.1690$ & This work \\ \hline
$E_b^{(a)}$ & $-0.2096$ & This work \\ \hline
$E_b^{(b)}$ & $-0.1520$ & This work \\ \hline
$E_b^{(c)}$ & $-0.3079$ & This work \\ \hline
$E_b^{(d)}$ & $-0.2992$ & This work \\ \hline
$E_{b,\text{1nn}}^{\text{V-V}}$ & -0.0146 & This work$^{\eref{xc}}$ \\
\hline
$E_{b,\text{2nn}}^{\text{V-V}}$ &  0.3028 & This work$^{\eref{xc}}$ \\
\hline
$E_f^\text{AA}$ & $10.16$ & Ref.\ \cite{2015JNMGharaee} \\ \hline
$E_f^\text{AB}$ &   $9.49$ & Ref.\ \cite{2015JNMGharaee} \\ \hline
$E_{b,\text{1nn}}^{\text{AA-B}}$ & $-0.52$ & Ref.\ \cite{2015JNMGharaee} \\ \hline 
$E_{b,\text{1nn}}^{\text{AB-B}}$ & $-0.53$ & Ref.\ \cite{setyawan_report} \\ \hline 
$E_{b,\text{1nn}}^{\text{AA-AA}}$ & $-2.12$ & Ref. \cite{2010JNMBecquart} \\ \hline
$E_{b,\text{1nn}}^{\text{AA-AB}}$ & $-2.12$ & Assumed $^{\eref{tabf}}$ \\ 
\hline
$E_{b,\text{1nn}}^{\text{AB-AB}}$ & $-3.2$  & Ref.\ \cite{2016JAPGharaee} \\ \hline
$E_m^{\text{V}\rightarrow \text{A}}$ (A) & $1.623$ & This work \\ \hline
$E_m^{\text{V}\rightarrow \text{B}}$ (A) & $1.651$ & This work \\ \hline
$E_m^{\text{V}\rightarrow \text{A(1)}}$ (Fig.\ \ref{fig_bvconf}(c)) & $1.7151$ & This work \\ \hline
$E_m^{\text{V}\rightarrow \text{A(2)}}$ (Fig.\ \ref{fig_bvconf}(c)) & $1.6378$ & This work \\ \hline
$E_m^{\text{V}\rightarrow \text{B(3)}}$ (Fig.\ \ref{fig_bvconf}(c)) & $1.577$ & This work \\ \hline
$E_m^{\text{V}\rightarrow \text{A}}$ (V) & $1.623$ & This work \\ \hline
$E_m^{\text{V}\rightarrow \text{B}}$ (V) & $1.651$ & This work \\ \hline
	\end{tabular}
	
\end{table}
\footnote{\label{xc}With xc-energy correction from Ref.\ \cite{2011JNMKato}} 

\begin{figure}[h]
	\centering
	\includegraphics[width=\columnwidth]{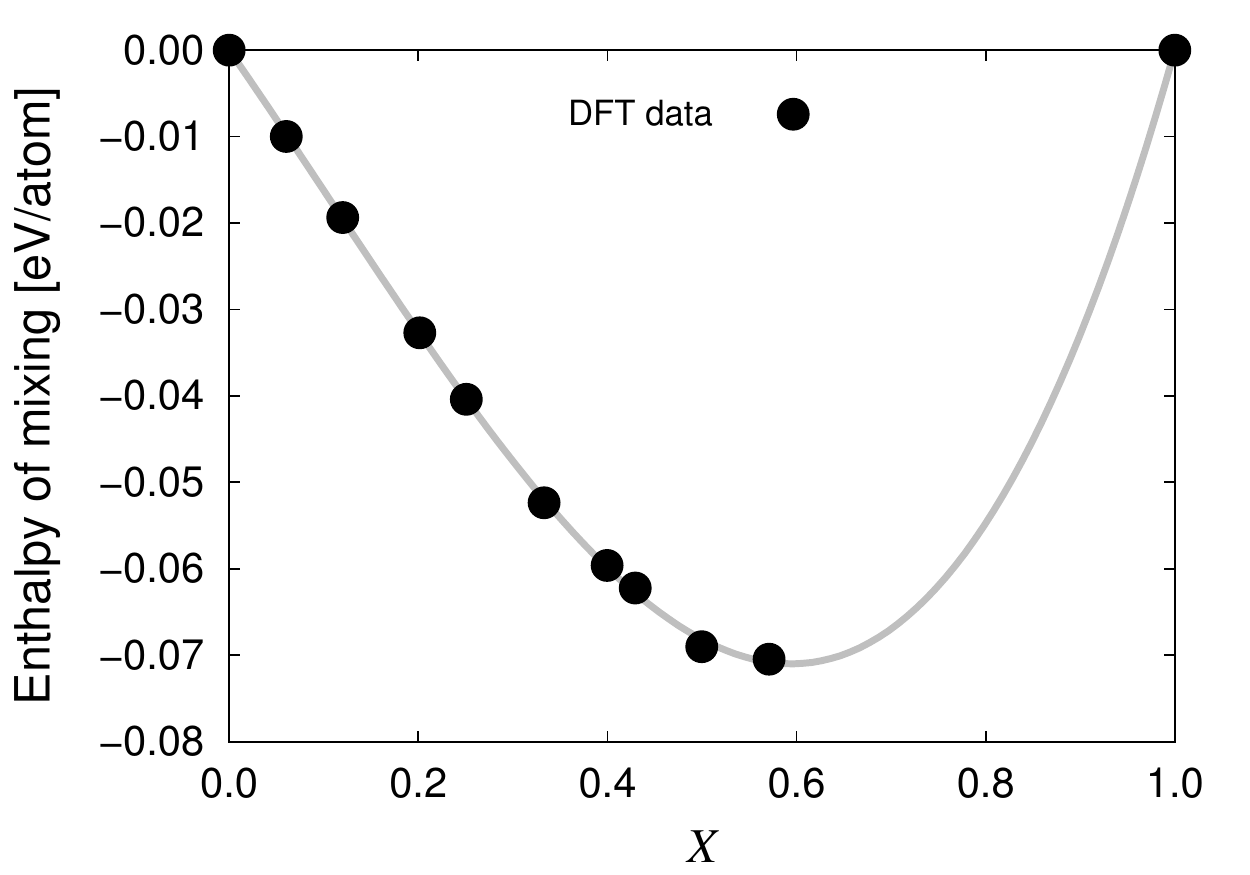}
	\caption{Enthalpy of mixing as a function of solute concentration from ref.\ \cite{2016JAPGharaee} and $3^{\rm rd}$-degree polynomial fit.}
	\label{fig_emix}
\end{figure}

The coefficient for the A-B bond is obtained from the enthalpy of mixing of W-Re, $\Delta H^{\rm mix}$, which can be written within the Bragg-Williams approximation \cite{1934BraggI, 1935BraggII, 1935BraggIII} as:
\begin{equation} \label{eq_bw}
    \begin{aligned}
\Delta H^{\rm mix}=   \: & \frac{z_1}{2}\left[(1-X)\epsilon^{(1)}_{\text{A-A}} + X\epsilon^{(1)}_{\text{B-B}} + 2x(1-x)\Omega^{(1)}_s \right] \\
        	    +\: & \frac{z_2}{2}\left[(1-X)\epsilon^{(2)}_{\text{A-A}} + X\epsilon^{(2)}_{\text{B-B}} + 2X(1-X)\Omega^{(2)}_s \right]
    \end{aligned}
\end{equation}
where $X$ is the global solute concentration, and $\Omega_s$ is the \emph{heat of solution}, defined as: 
\begin{align}
 \label{eq_bw1}
\Omega_s^{(1)}= \epsilon_{\text{A-B}}^{(1)} - \frac{1}{2} \left( \epsilon_{\text{A-A}}^{(1)} + \epsilon_{\text{B-B}}^{(1)} \right) \\
\Omega_s^{(2)}= \epsilon_{\text{A-B}}^{(2)} - \frac{1}{2} \left( \epsilon_{\text{A-A}}^{(2)} + \epsilon_{\text{B-B}}^{(2)} \right)  
\end{align}
Combining eqs.\ \eref{eq_coh} and \eref{eq_bw}, $\Delta E^{mix}$ can be expressed as:
\begin{equation} \label{eq_bw2}
    X (1-X) \Omega_s^*= \Delta H^{\rm mix} + (1-X) E_{coh}^\text{A} + X E_{coh}^\text{B} 
\end{equation}
where $\Omega_s^*= z_1 \Omega_s^{(1)} + z_2 \Omega_s^{(2)}$. 
To ascertain the dependence on the solute concentration of the heat of solution, we fit the l.h.s.\ of eq.\ \eref{eq_bw2} to the data points for the mixing enthalpies as a function of $X$ calculated in our previous work \cite{2016JAPGharaee}. The best fit, shown in Figure \ref{fig_emix}, is achieved when $\Omega_s^*$ is expressed a linear function of the concentration: 
$$\Omega_s^*= w_0 + w_1X$$
with $w_0=-0.1571$ and $w_1=-0.2311$. The negative values of $w_0$ and $w_1$ suggest a strong tendency towards ordering, which becomes larger as the solute concentration increases. Combining eqs.\ \eref{eq_e2e1}, \eref{eq_coh}, \eref{eq_bw1}, and \eref{eq_bw2}, one can obtain the values of $\Omega_s^{(1)}$, $\Omega_s^{(2)}$, $\epsilon_{\text{A-B}}^{(1)}$, and $\epsilon_{\text{A-B}}^{(2)}$.
A non-constant $\Omega_s^*$ effectively implies that $\epsilon_{\text{A-B}}$ is also a function of the concentration.  
Moreover, to reflect local composition variations in the W-Re alloys, we make the assumption that the dependence of $\epsilon_{\text{A-B}}^{(1)}$ and $\epsilon_{\text{A-B}}^{(2)}$ on $X$ can be transferred to the local environment of each atom, such that both bond energy coefficients are functions of the local composition, which we term $x$, and must be computed on the fly for each solute atom in the system. 

\subsubsection{Vacancy parameters}
\begin{figure}[h]
	\centering
		\begin{tikzpicture}		
	\draw[dashed, line width=0.3mm] (-0.5,  0.5, -0.5) -- (-0.5, -0.5, -0.5);
	\draw[dashed, line width=0.3mm] (-0.5, -0.5, -0.5) -- ( 0.5, -0.5, -0.5);
	\draw[dashed, line width=0.3mm] (-0.5, -0.5, -0.5) -- (-0.5, -0.5,  0.5);
	\draw[line width=0.3mm] (-0.5,  0.5, -0.5) -- ( 0.5,  0.5, -0.5);
	\draw[line width=0.3mm] (-0.5,  0.5, -0.5) -- (-0.5,  0.5,  0.5);
	\draw[line width=0.3mm] (-0.5,  0.5,  0.5) -- ( 0.5,  0.5,  0.5);
	\draw[line width=0.3mm] ( 0.5,  0.5,  0.5) -- ( 0.5,  0.5, -0.5);
	\draw[line width=0.3mm] (-0.5,  0.5,  0.5) -- (-0.5, -0.5,  0.5);
	\draw[line width=0.3mm] (-0.5, -0.5,  0.5) -- ( 0.5, -0.5,  0.5);
	\draw[line width=0.3mm] ( 0.5, -0.5,  0.5) -- ( 0.5,  0.5,  0.5);
	\draw[line width=0.3mm] ( 0.5,  0.5, -0.5) -- ( 0.5, -0.5, -0.5);
	\draw[line width=0.3mm] ( 0.5, -0.5,  0.5) -- ( 0.5, -0.5, -0.5);
	\shade[ball color= blue] ( 0,0,0) circle (0.2cm);
	\shade[ball color= red]   ( -0.5, 0.5, -0.5) circle (0.2cm);
\begin{scope}[shift={(1.5, 0)}]
	\draw[dashed, line width=0.3mm] (-0.5,  0.5, -0.5) -- (-0.5, -0.5, -0.5);
	\draw[dashed, line width=0.3mm] (-0.5, -0.5, -0.5) -- ( 0.5, -0.5, -0.5);
	\draw[dashed, line width=0.3mm] (-0.5, -0.5, -0.5) -- (-0.5, -0.5,  0.5);
	\draw[line width=0.3mm] (-0.5,  0.5, -0.5) -- ( 0.5,  0.5, -0.5);
	\draw[line width=0.3mm] (-0.5,  0.5, -0.5) -- (-0.5,  0.5,  0.5);
	\draw[line width=0.3mm] (-0.5,  0.5,  0.5) -- ( 0.5,  0.5,  0.5);
	\draw[line width=0.3mm] ( 0.5,  0.5,  0.5) -- ( 0.5,  0.5, -0.5);
	\draw[line width=0.3mm] (-0.5,  0.5,  0.5) -- (-0.5, -0.5,  0.5);
	\draw[line width=0.3mm] (-0.5, -0.5,  0.5) -- ( 0.5, -0.5,  0.5);
	\draw[line width=0.3mm] ( 0.5, -0.5,  0.5) -- ( 0.5,  0.5,  0.5);
	\draw[line width=0.3mm] ( 0.5,  0.5, -0.5) -- ( 0.5, -0.5, -0.5);
	\draw[line width=0.3mm] ( 0.5, -0.5,  0.5) -- ( 0.5, -0.5, -0.5);
	\draw[line width=0.3mm] (0,0,0) -- ( 0, 1, 0);
	\shade[ball color= blue] ( 0,0,0) circle (0.2cm);
	\shade[ball color= red]   ( 0, 1, 0) circle (0.2cm);
\end{scope}	
\begin{scope}[shift={(3.0, 0)}]
	\draw[dashed, line width=0.3mm] (-0.5,  0.5, -0.5) -- (-0.5, -0.5, -0.5);
	\draw[dashed, line width=0.3mm] (-0.5, -0.5, -0.5) -- ( 0.5, -0.5, -0.5);
	\draw[dashed, line width=0.3mm] (-0.5, -0.5, -0.5) -- (-0.5, -0.5,  0.5);
	\draw[line width=0.3mm] (-0.5,  0.5, -0.5) -- ( 0.5,  0.5, -0.5);
	\draw[line width=0.3mm] (-0.5,  0.5, -0.5) -- (-0.5,  0.5,  0.5);
	\draw[line width=0.3mm] (-0.5,  0.5,  0.5) -- ( 0.5,  0.5,  0.5);
	\draw[line width=0.3mm] ( 0.5,  0.5,  0.5) -- ( 0.5,  0.5, -0.5);
	\draw[line width=0.3mm] (-0.5,  0.5,  0.5) -- (-0.5, -0.5,  0.5);
	\draw[line width=0.3mm] (-0.5, -0.5,  0.5) -- ( 0.5, -0.5,  0.5);
	\draw[line width=0.3mm] ( 0.5, -0.5,  0.5) -- ( 0.5,  0.5,  0.5);
	\draw[line width=0.3mm] ( 0.5,  0.5, -0.5) -- ( 0.5, -0.5, -0.5);
	\draw[line width=0.3mm] ( 0.5, -0.5,  0.5) -- ( 0.5, -0.5, -0.5);
	\shade[ball color= blue] ( 0,0,0) circle (0.2cm);
	\shade[ball color= red]   ( -0.5, 0.5, -0.5) circle (0.2cm); \node at ( -0.5, 0.5, -0.5) {3};
	\shade[ball color= red]   (  0.5, 0.5, -0.5) circle (0.2cm);
	\shade[ball color= green]   ( 0.5, 0.5, 0.5) circle (0.2cm); \node at ( 0.5, 0.5, 0.5) {1};
	\shade[ball color= green]   ( 0.5, -0.5, 0.5) circle (0.2cm); \node at ( 0.5, -0.5, 0.5) {2};
\end{scope}
\begin{scope}[shift={(4.5, 0)}]
	\draw[dashed, line width=0.3mm] (-0.5,  0.5, -0.5) -- (-0.5, -0.5, -0.5);
	\draw[dashed, line width=0.3mm] (-0.5, -0.5, -0.5) -- ( 0.5, -0.5, -0.5);
	\draw[dashed, line width=0.3mm] (-0.5, -0.5, -0.5) -- (-0.5, -0.5,  0.5);
	\draw[line width=0.3mm] (-0.5,  0.5, -0.5) -- ( 0.5,  0.5, -0.5);
	\draw[line width=0.3mm] (-0.5,  0.5, -0.5) -- (-0.5,  0.5,  0.5);
	\draw[line width=0.3mm] (-0.5,  0.5,  0.5) -- ( 0.5,  0.5,  0.5);
	\draw[line width=0.3mm] ( 0.5,  0.5,  0.5) -- ( 0.5,  0.5, -0.5);
	\draw[line width=0.3mm] (-0.5,  0.5,  0.5) -- (-0.5, -0.5,  0.5);
	\draw[line width=0.3mm] (-0.5, -0.5,  0.5) -- ( 0.5, -0.5,  0.5);
	\draw[line width=0.3mm] ( 0.5, -0.5,  0.5) -- ( 0.5,  0.5,  0.5);
	\draw[line width=0.3mm] ( 0.5,  0.5, -0.5) -- ( 0.5, -0.5, -0.5);
	\draw[line width=0.3mm] ( 0.5, -0.5,  0.5) -- ( 0.5, -0.5, -0.5);
	\draw[line width=0.3mm] (0,0,0) -- ( 0,  1, 0);
	\draw[line width=0.3mm] (0,0,0) -- ( 0, -1, 0);
	\shade[ball color= blue] ( 0,0,0) circle (0.2cm);
	\shade[ball color= red]   ( 0,  1, 0) circle (0.2cm);
	\shade[ball color= red]   ( 0, -1, 0) circle (0.2cm);
\end{scope}
	\node[font=\large] at (   0, -1.5) {(a)};
	\node[font=\large] at (1.5, -1.5) {(b)};
	\node[font=\large] at (3.0, -1.5) {(c)};
	\node[font=\large] at (4.5, -1.5) {(d)};
		\end{tikzpicture}
		
	\caption{Configurations of V-Re clusters used to extract bond energy coefficients $\epsilon_{\text{A-V}}$ and $\epsilon_{\text{B-V}}$. Blue spheres represent vacancies, red spheres represents Re atoms. All other lattice sites are occupied by A atoms, which are omitted for clarity. Green spheres indicate the various equivalent sites for atoms to exchange positions with the vacancy\label{fig_bvconf}}
\end{figure}
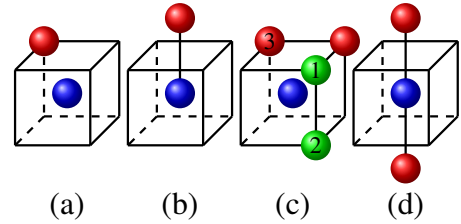

The vacancy bond coefficients are $\epsilon_{\text{A-V}}$, $\epsilon_{\text{B-V}}$, and $\epsilon_{\text{V-V}}$.  $\epsilon_{\text{A-V}}$ can be readily obtained from the value of the vacancy formation energy:
\begin{equation}
	E_f^\text{V}= E_{coh}^\text{A} -z_1\epsilon_{\text{A-V}}^{(1)}-z_2\epsilon_{\text{A-V}}^{(2)}
\end{equation}
where $E_f^{\text{V}}$ is the vacancy formation energy in pure W (given in Table \ref{dftt}). $\epsilon_{\text{B-V}}$ can be obtained from the binding energies of V-Re configurations. The binding energy of a configuration involving $m$ solute atoms and $n$ vacancies is defined as:
\begin{equation} \label{eq_bind}
E_b^{\text{B}_m \text{V}_n}= E_f^{\text{B}_m \text{V}_n} - m E_f^{\text{B}} - n E_f^{\text{V}}
\end{equation} 
where the $E_f$ are the respective formation energies of each structure. In this work, binding energies for the four vacancy-solute configurations shown in Fig.\ \eref{fig_bvconf} have been calculated (cf.\ Table \ref{dftt}). One can now rewrite eq.\ \eref{eq_bind} as a function of the B-V bond coefficients $\epsilon_{\text{B-V}}^{(1)}$ and $\epsilon_{\text{B-V}}^{(2)}$ for each one of the configurations in the figure:
\begin{subequations}
	\begin{align}
& E_{b}^{(a)} = \epsilon_{\text{B-V}}^{(1)}+ \epsilon_{\text{A-A}}^{(1)} - \epsilon_{\text{A-B}}^{(1)} - \epsilon_{\text{A-V}}^{(1)} \\
& E_{b}^{(b)} = \epsilon_{\text{B-V}}^{(2)}+ \epsilon_{\text{A-A}}^{(2)} - \epsilon_{\text{A-B}}^{(2)}- \epsilon_{\text{A-V}}^{(2)} \\
&		\begin{aligned}
E_{b}^{(c)}= & 2\epsilon_{\text{B-V}}^{(1)} + \epsilon_{\text{B-B}}^{(2)} + 2\epsilon_{\text{A-A}}^{(1)} + \epsilon_{\text{A-A}}^{(2)} - 2\epsilon_{\text{A-V}}^{(1)} -2\epsilon_{\text{A-B}}^{(1)} \\
& - 2\epsilon_{\text{A-B}}^{(2)} + 14 \Delta \epsilon_{\text{A-B}}^{(1)} + 10 \Delta \epsilon_{\text{A-B}}^{(2)}
		\end{aligned} \\
& E_{b}^{(d)}= 2\epsilon_{\text{B-V}}^{(2)}+2\epsilon_{\text{A-A}}^{(2)} - 2\epsilon_{\text{A-B}}^{(2)}- 2\epsilon_{\text{A-V}}^{(2)}	
	\end{align}
\end{subequations}
where $\Delta \epsilon_\text{A-B}^{(m)}$ is the change in $\epsilon_\text{A-B}^{(m)}$ due to the local solute concentration change resulting from the vacancy jump. 

To define the dependence on $x$ of $\epsilon_{\text{B-V}}^{(1)}$, we must consider two factors. First, our DFT calculations show that $\epsilon_{\text{A-V}}^{(1)}>\epsilon_{\text{B-V}}^{(1)}$. Second, $x$ of $\epsilon_{\text{B-V}}^{(1)}$ is seen to increase with local concentration. Both of these conditions are satisfied by assuming a dependence such as $\epsilon_{\text{B-V}}^{(1)}(x)= \epsilon_{\text{A-V}}^{(1)}-ax^{-1}$, where $a$ is a fitting constant. As well, $\epsilon_{\text{B-V}}^{(2)}$ is seen to independently increase with concentration, such that $\epsilon_{\text{B-V}}^{(2)}(x)= bx+c$, where $b$ and $c$ are fitting parameters

$\epsilon_{\text{V-V}}$ can be readily calculated by considering the binding energy of a di-vacancy: 
\begin{gather}
E_{b,\text{1nn}}^{\text{V-V}}= \epsilon_{\text{A-A}}^{(1)} + \epsilon_{\text{V-V}}^{(1)} - 2 \epsilon_{\text{A-V}}^{(1)}  \\
E_{b,\text{2nn}}^{\text{V-V}}= \epsilon_{\text{A-A}}^{(2)} + \epsilon_{\text{V-V}}^{(2)} - 2 \epsilon_{\text{A-V}}^{(2)}
\end{gather}
It is interesting to note that, in accordance with several other studies \cite{Becquart200723,ventelon2012ab,PhysRevB.84.104115,oda2014,setyawan_report}, $E_{b,\text{2nn}}^{\text{V-V}}$ takes a positive value (cf.\ Table \ref{dftt}), indicating repulsion between vacancies that are at 2nn distances of each other.

\subsubsection{Interstitial defect parameters}
The interstitial bond coefficients include $\epsilon_{\text{AA-A}}$, $\epsilon_{\text{AB-A}}$, $\epsilon_{\text{AA-B}}$, $\epsilon_{\text{AB-B}}$, $\epsilon_{\text{AA-AA}}$, $\epsilon_{\text{AA-AB}}$, and $\epsilon_{\text{AB-AB}}$. $\epsilon_{\text{AA-A}}$ and $\epsilon_{\text{AB-A}}$ are calculated directly from the formation energies of SIAs and mixed dumbbells:
\begin{gather}
E_{f}^{\text{AA}}= -4 \epsilon_{\text{A-A}}^{(1)} -3 \epsilon_{\text{A-A}}^{(2)} + 8 \epsilon_{\text{AA-A}}^{(1)} + 6 \epsilon_{\text{AA-A}}^{(2)} \\
E_{f}^{\text{AB}}= -4 \epsilon_{\text{A-A}}^{(1)}  -3 \epsilon_{\text{A-A}}^{(2)} + 8 \epsilon_{\text{AB-A}}^{(1)} + 6 \epsilon_{\text{AB-A}}^{(2)}
\end{gather}
The other bond coefficients are obtained from various binding energies:
\begin{align}
E_{b,\text{1nn}}^{\text{AA-B}} &= \epsilon_{\text{AA-B}}^{(1)} + \epsilon_{\text{A-A}}^{(1)} - \epsilon_{\text{AA-A}}^{(1)} - \epsilon_{\text{A-B}}^{(1)} \\
E_{b,\text{1nn}}^{\text{AB-B}} &= \epsilon_{\text{AB-B}}^{(1)} + \epsilon_{\text{A-A}}^{(1)} - \epsilon_{\text{AB-A}}^{(1)} - \epsilon_{\text{A-B}}^{(1)} \\
E_{b,\text{1nn}}^{\text{AA-AA}} &= \epsilon_{\text{AA-AA}}^{(1)} + \epsilon_{\text{A-A}}^{(1)} - 2 \epsilon_{\text{AA-A}}^{(1)} \\
E_{b,\text{1nn}}^{\text{AA-AB}} &= \epsilon_{\text{AA-AB}}^{(1)} + \epsilon_{\text{A-A}}^{(1)} - \epsilon_{\text{AA-A}}^{(1)} - \epsilon_{\text{AB-A}}^{(1)} \\
E_{b,\text{1nn}}^{\text{AB-AB}} &= \epsilon_{\text{AB-AB}}^{(1)} + \epsilon_{\text{A-A}}^{(1)} - 2 \epsilon_{\text{AB-A}}^{(1)}
\end{align}
These formation and binding energies are all taken from the literature\footnote{\label{tabf}The only exception being the binding energy between an AA and an AB interstitial, which is assumed to be equal to the binding energy between two AA.}. 

All the bond energy coefficients, the equation used for their calculation, and the source of the numbers are compiled in Table \ref{tab_coeff}.

\begin{table*}[htbp]
\begin{minipage}{\linewidth}
\centering
	\caption{Bond energy coefficients with the equation used for their calculation, and the literature source. $x$ is the local solute concentration}
	\label{tab_coeff}
\begin{tabular}{| c | c | c | c |}
	\hline
$\epsilon_{\text{A-A}}^{(1)}$ & $-1.5815$ & cohesive energy, eq.\ \eref{eq_e2e1} & This work \\ \hline
$\epsilon_{\text{A-A}}^{(2)}$ & $-0.6672$ & cohesive energy, eq.\ \eref{eq_e2e1} & This work \\ \hline
$\epsilon_{\text{B-B}}^{(1)}$ & $-1.4067$ & cohesive energy, eq.\ \eref{eq_e2e1} & This work \\ \hline 
$\epsilon_{\text{B-B}}^{(2)}$ & $-0.5935$ & cohesive energy, eq.\ \eref{eq_e2e1} & This work \\ \hline
$\epsilon_{\text{A-B}}^{(1)}$ & $-1.5090 - 0.0219 x$ & mixing energy & Ref. \cite{2016JAPGharaee} 
\\ \hline
$\epsilon_{\text{A-B}}^{(2)}$ & $-0.6366 - 0.0092 x$ & eq. \eref{eq_e2e1} & Ref. \cite{2016JAPGharaee}  
\\ \hline
$\epsilon_{\text{A-V}}^{(1)}$ & $-0.4898$ & formation energy, eq. \eref{eq_e2e1} & This work \\ \hline
$\epsilon_{\text{A-V}}^{(2)}$ & $-0.2067$ & formation energy, eq. \eref{eq_e2e1} & This work \\ \hline
$\epsilon_{\text{B-V}}^{(1)}$ & $-0.4898-0.009432/x$ & formation energy fitted to $\epsilon_{B-V}^{(1)}= a+b/x$ & This work 	\\ \hline
$\epsilon_{\text{B-V}}^{(2)}$ & $-0.3311+0.036x$ & formation energy fitted to $\epsilon_{B-V}^{(1)}= a+bx$ & This work 	\\ \hline
$\epsilon_{\text{V-V}}^{(1)}$ & $0.5873$ & 1nn binding energy & This work \\ \hline
$\epsilon_{\text{V-V}}^{(2)}$ & $0.5566$ & 2nn binding energy & This work \\ \hline
$\epsilon_{\text{AA-A}}^{(1)}$ & $0.1740$ & formation energy, eq.\ \eref{eq_e2e1} & Ref. \cite{2015JNMGharaee}
\\ \hline
$\epsilon_{\text{AA-A}}^{(2)}$ & $0.0734$ & formation energy, eq.\ \eref{eq_e2e1} & Ref. \cite{2015JNMGharaee}
\\ \hline
$\epsilon_{\text{AB-A}}^{(1)}$ & $0.1104$ & formation energy, eq.\ \eref{eq_e2e1} & Ref. \cite{2015JNMGharaee}
\\ \hline
$\epsilon_{\text{AB-A}}^{(2)}$ & $0.0466$ & formation energy, eq.\ \eref{eq_e2e1} & Ref. \cite{2015JNMGharaee}
\\ \hline
$\epsilon_{\text{AA-B}}^{(1)}$ & $-0.2750$ & binding energy & Ref.\ \cite{2015JNMGharaee}			\\ \hline  
$\epsilon_{\text{AA-B}}^{(2)}$ & $-0.1160$ & eq.\ \eref{eq_e2e1} & Ref.\ \cite{2015JNMGharaee}		\\ \hline
$\epsilon_{\text{AB-B}}^{(1)}$ & $-0.3486$ & binding energy & Ref.\ \cite{setyawan_report}				\\ \hline
$\epsilon_{\text{AB-B}}^{(2)}$ & $-0.1470$ & eq.\ \eref{eq_e2e1} & Ref.\ \cite{setyawan_report}			\\ \hline
$\epsilon_{\text{AA-AA}}^{(1)}$ & $-0.1905$ & binding energy & Ref. \cite{2010JNMBecquart}		\\ \hline 
$\epsilon_{\text{AA-AA}}^{(2)}$ & $-0.0804$ & eq.\ \eref{eq_e2e1} & Ref. \cite{2010JNMBecquart}		\\ \hline
$\epsilon_{\text{AA-AB}}^{(1)}$ & $-0.2505$ & binding energy & Assumed $^{\eref{tabf}}$				\\ \hline
$\epsilon_{\text{AA-AB}}^{(2)}$ & $-0.1057$ & eq.\ \eref{eq_e2e1} & Assumed $^{\eref{tabf}}$		\\ \hline
$\epsilon_{\text{AB-AB}}^{(1)}$ & $-1.3977$ & binding energy & Ref.\ \cite{2016JAPGharaee}		\\ \hline 
$\epsilon_{\text{AB-AB}}^{(2)}$ & $-0.5897$ & eq.\ \eref{eq_e2e1} & Ref.\ \cite{2016JAPGharaee}		\\ \hline
	\end{tabular}
\end{minipage}
\end{table*}

\subsubsection{Migration parameters} \label{sec_mig}

The attempt frequency ($\nu$ in eq.\ \eref{eq_rate}) used for vacancy jumps in this work is set to be equal to Debye frequency of W, or $6.5 \times 10^{12}$ Hz \cite{2008JPCMGilbert}, while for interstitials we use a value of $1.5\times10^{12}$ Hz \cite{2013JNMZhou}.

From eq.\ \eref{eq_sp}, there are six different saddle-point bond coefficients: $\epsilon_\text{A-A}^{sp(m)}$, $\epsilon_\text{A-B}^{sp(m)}$, $\epsilon_\text{A-V}^{sp(m)}$, $\epsilon_\text{B-A}^{sp(m)}$, $\epsilon_\text{B-B}^{sp(m)}$, and $\epsilon_\text{B-V}^{sp(m)}$, where $m$= 1nn, 2nn. In this notation, $\epsilon_{\alpha \text{-} \beta}^{sp(m)}$ represents the energy of the bond between the atom at the saddle point $\alpha$ and its closest lattice neighbor $\beta$. This means $\epsilon_{\alpha \text{-} \beta}^{sp(m)}\ne\epsilon_{\beta \text{-} \alpha}^{sp(m)}$. 

The saddle-point bond coefficients connected to a lattice atom A (W atom), $\epsilon_{\alpha \text{-A}}^{sp(m)}$, can be calculated as:
\begin{equation} \label{eq_spA}
z_1^{sp} \epsilon_{\alpha \text{-A}}^{sp(1)} + z_2^{sp} \epsilon_{\alpha \text{-A}}^{sp(2)}=	E_m + \sum_{n, q} \epsilon_{X\text{-}q}^{(n)} + \sum_{n, r \neq X} \epsilon_{V\text{-}r}^{(n)}
\end{equation}
where $z_1^{sp}$ and $z_2^{sp}$ are the numbers of $1^{\rm st}$- and $2^{\rm nd}$ nearest neighbor of an atom at the  saddle point, which are both equal to 6 for the bcc lattice, and $E_m$ is the migration energy. The term $\Delta E_{ij}^{\rm non\text{-}broken}$ in eq.\ \eref{eq_sp} is zero here since no solute concentration change is involved in an A-atom jump. $\epsilon_{\alpha \text{-A}}^{sp(2)}$ is obtained from $\epsilon_{\alpha \text{-A}}^{sp(1)}$ using eq.\ \eref{eq_e2e1}. Vacancy bonds are calculated in a similar manner. 

To calculate the saddle-point bond coefficients pertaining to B (Re) atoms, $\epsilon_{\alpha \text{-B}}^{sp(m)}$, one must consider local solute concentration changes. To this end, we resort to the configurations shown in Fig.\ \ref{fig_bvconf}(c). The A-B saddle-point coefficients $\epsilon_{\text{A-B}}^{sp(m)}$ are obtained from A-atom jumps, labeled `1' and `2' in Fig.\ \ref{fig_bvconf}(c), into the vacant site. 
The B-B saddle-point coefficient $\epsilon_{\text{B-B}}^{sp(1)}$ is computed assuming a B-atom (labeled `3' in the figure) jump into the vacancy. Equation \eref{eq_e2e1} is then used to obtain the 2nn coefficients. All the necessary DFT calculations to calculate the saddle-point bond coefficients were performed as part of the present work, and are given in Table \eref{tab_spcoeff}.
\begin{table}[h]
	\caption{Saddle-point bond energy coefficients for vacancy jumps (in eV).}
	\label{tab_spcoeff}
	\centering
	\begin{tabular}{| c | c | c | c |}
	\hline
$\epsilon_{\text{A-A}}^{sp(1)}$ & $-2.5975$ & $\epsilon_{\text{A-A}}^{sp(2)}$ & $-0.5041$ 	\\ \hline
$\epsilon_{\text{A-B}}^{sp(1)}$ & $-2.6451$ & $\epsilon_{\text{A-B}}^{sp(2)}$ & $-0.5532$ 	\\ \hline
$\epsilon_{\text{A-V}}^{sp(1)}$ & $0.5465$ & $\epsilon_{\text{A-V}}^{sp(2)}$ & $0.1060$ 	\\ \hline
$\epsilon_{\text{B-A}}^{sp(1)}$ & $-2.5188$ & $\epsilon_{\text{B-A}}^{sp(2)}$ & $-0.4888$ 	\\ \hline
$\epsilon_{\text{B-B}}^{sp(1)}$ & $-2.5417$ & $\epsilon_{\text{B-B}}^{sp(2)}$ & $-0.4943$ 	\\ \hline
$\epsilon_{\text{B-V}}^{sp(1)}$ & $0.2902$ & $\epsilon_{\text{B-V}}^{sp(2)}$ & $0.0563$ 	\\ \hline
	\end{tabular}
\end{table}

The migration energies of SIA and mixed-interstitials, the activation energy for SIA rotation, as well as the correlation factors at different temperatures are taken from the literature, and listed in Table \eref{tab_mig}.
\begin{table}[tbh]
	\caption{Self-interstitial migration parameters. The jump distance for SIA migration is $\delta=a_0\sqrt{3}/2$.}
	\label{tab_mig}
	\centering
	\begin{tabular}{| c | c | c |}
	\hline
$E_m^{\text{AA}}$ & 0.003 & Ref. \cite{2014MSMSESuzudo} \\
	\hline
$E_r^{\text{AA}}$ & 0.43 & Ref. \cite{2014MSMSESuzudo} \\
	\hline
$E_m^{\text{AB}}$ & 0.12 & Ref. \cite{2016JAPGharaee} \\
	\hline
$f$ & $2.93 - 0.00055 T$ & Ref. \cite{2013JNMZhou} \\
	\hline
	\end{tabular}
\end{table}

\subsubsection{DFT calculations}
Density functional theory calculations were carried out using the projector augmented wave (PAW) method \cite{Blo94,KreJou99} as implemented in the Vienna {\it ab-initio} simulation package \cite{KreHaf93, KreHaf94, KreFur96a, KreFur96b}. Since interstitial configurations involve short interatomic distances ``hard'' PAW setups that include semi-core electron states were employed with a plane wave energy cutoff of 300 eV.

Exchange and correlation effects were described using the generalized gradient approximation \cite{PerBurErn96} while the occupation of electronic states was performed using the first order Methfessel-Paxton scheme with a smearing width of 0.2\,eV. The Brillouin zone was sampled using $5\times 5 \times 5$ $\vec{k}$-point grids. (A detailed discussion of the effect of different computational parameters on the results can be found in Ref.~\cite{2015JNMGharaee}). All structures were optimized allowing full relaxation of both ionic positions and cell shape with forces converged to below 10 meV/\AA. Migration barriers were computed using $4\times4\times4$ supercells and the climbing image-nudged elastic band method with three images \cite{HenUbeJon00}.

\section{Results} \label{sec:res}
\subsection{Structural phase diagrams}\label{struct}
\begin{figure}[h]
	\centering
	\includegraphics[width=\columnwidth]{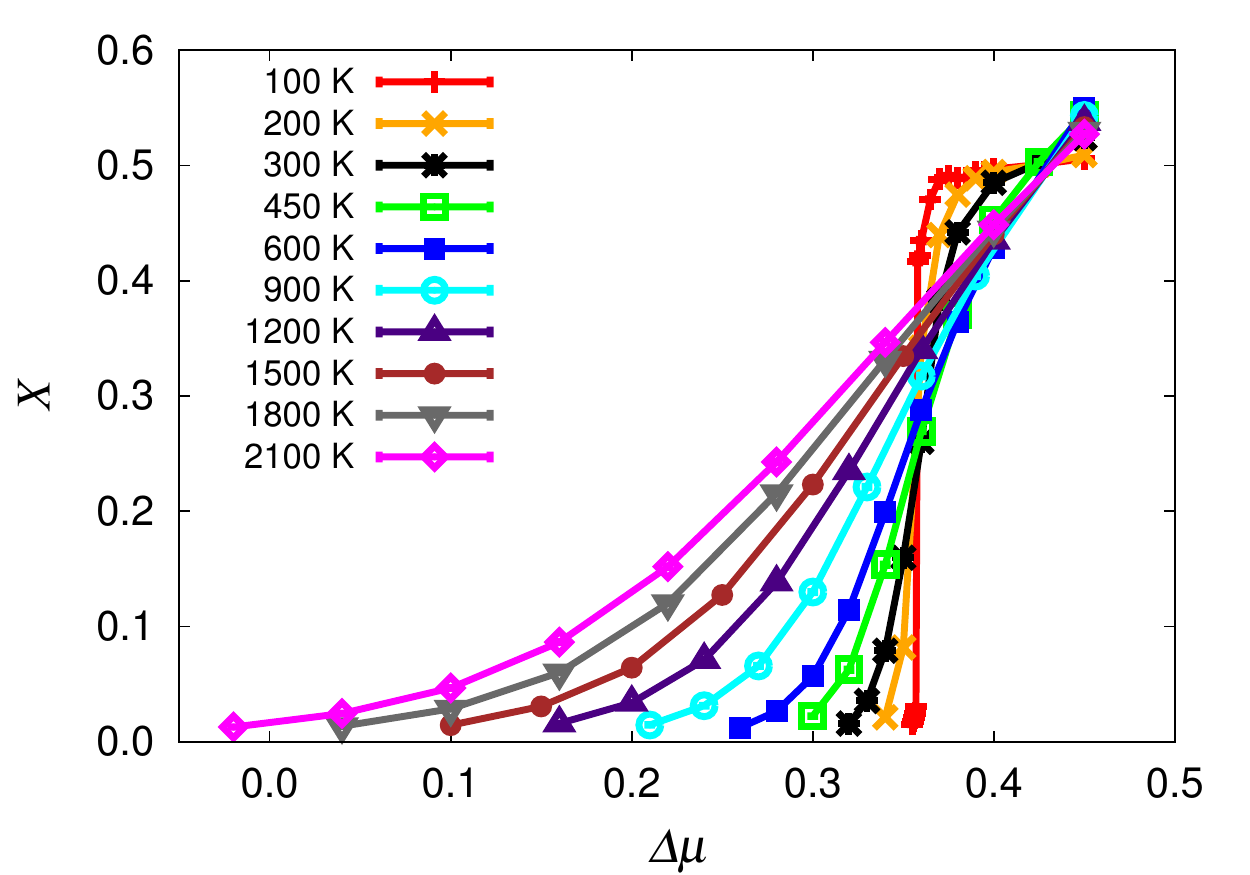}
	\caption{Solute composition $X$ as a function of chemical potential $\Delta\mu$ at different temperatures.}
	\label{fig_c000comp}
\end{figure}

Although our energy model includes thermodynamic information reflective of the phase stability of W-Re alloys, the model consists of a rigid lattice with bcc structure and is thus suitable only for a given, well-defined, concentration range. Our DFT calculations yield bond energies that are consistent with stable solid solutions from zero to approximately 40\% at.\ Re \cite{2016JAPGharaee}. This is confirmed by way of sgMC simulations performed as a function of composition and temperature in $64\times64\times64$ computational cells. Figure \ref{fig_c000comp} shows the set of stable compositions obtained as a function of the chemical potential for several temperatures.
The figure shows a clear jump in the Re concentration at a temperature of approximately 100 K. This is typically indicative of a phase boundary, as two distinct phases characterized by widely different solute concentrations seem to coexist at the same temperature and chemical potential. This may be indicative of a miscibility gap between Re concentrations of a few percent and approximately 50 at.\%, {\it i.e.} beyond the thermodynamic validity of our rigid lattice model. To further characterize the configurations obtained, we calculate the short-range order (SRO) of the configurations obtained according to the \emph{Warren-Cowley} parameter \cite{PhysRev.77.669}:
\begin{equation}
	\eta=N_B^{-1}\sum_i^{N_B}{\left(1-\frac{x_i({\rm A})}{1-X}\right)}
\end{equation}
which gives the SRO parameter $\eta$ of Re atoms w.r.t.\ matrix W atoms, with $x_i({\rm A})$ being the fraction of A atoms surrounding each solute atom $i$. The sum extends to all B atoms in the system.

\begin{figure}[h]
	\centering
	\includegraphics[width=\columnwidth]{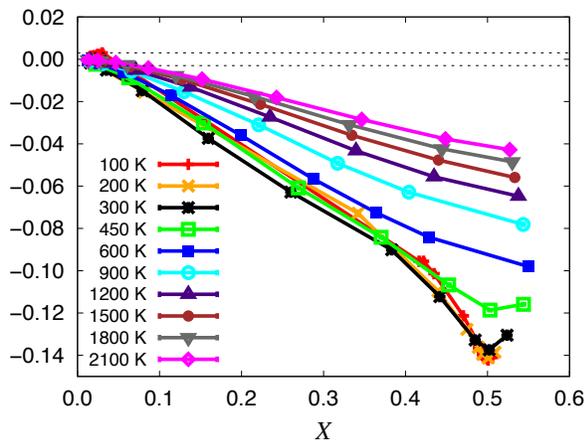}
	\caption{Short range order parameter $\eta$ as a function of global solute composition $X$ at different temperatures. The dashed line indicate the SRO interval caused by normal concentration fluctuations during the generation of atomistic samples.}
	\label{fig_c000sro}
\end{figure}

\begin{figure}[h]
	\centering
	\includegraphics[width=\columnwidth]{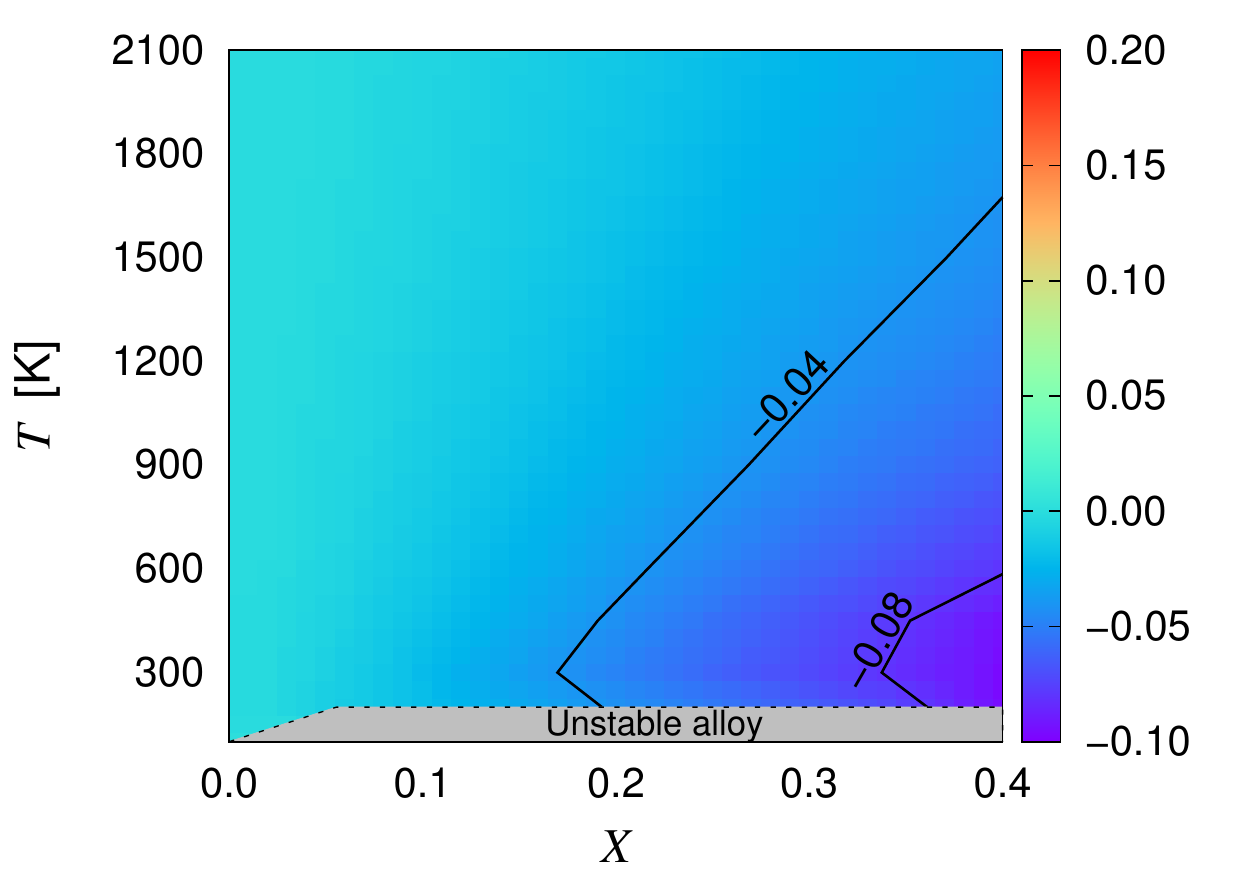}
	\caption{Structural phase diagram showing regions of changing SRO. The dashed lines are the limits of applicability of the rigid bcc lattice model. The system displays slightly negative SRO throughout the entire temperature-concentration space, indicating a preference to be in a solid solution state.}
	\label{fig_c000phase}
\end{figure}

According to this definition, $\eta>0$ implies phase separation, $\eta=0$ represents an ideal solid solution, and $\eta<0$ indicates ordering. However, the SRO parameter of a random solution has a range of $-0.003$ to 0.003 regardless of solute composition due to the random occurrence of dimers, trimers, etc. Figure \eref{fig_c000sro} shows the equilibrium SRO as a function of $X$ for several temperatures. 
As the figure shows, the SRO parameter is near zero for dilute systems, and gradually becomes negative as the concentration increases. Based on the figure we conclude that equilibrium W-Re systems with up to $\approx40$ at.\% solute content are consistent with random solid solutions with a weak tendency to ordering at higher concentrations. The corresponding $T$-$X$ phase diagram is provided in Figure \ref{fig_c000phase}.

\subsubsection{Effect of vacancies on phase diagram} \label{sec_defect}
\begin{figure}[h]
\centering
	\begin{minipage}{0.32\textwidth}
		\centering
		\subfloat[$C_v=0.01$ at.\%]{\includegraphics[width=0.99\columnwidth]{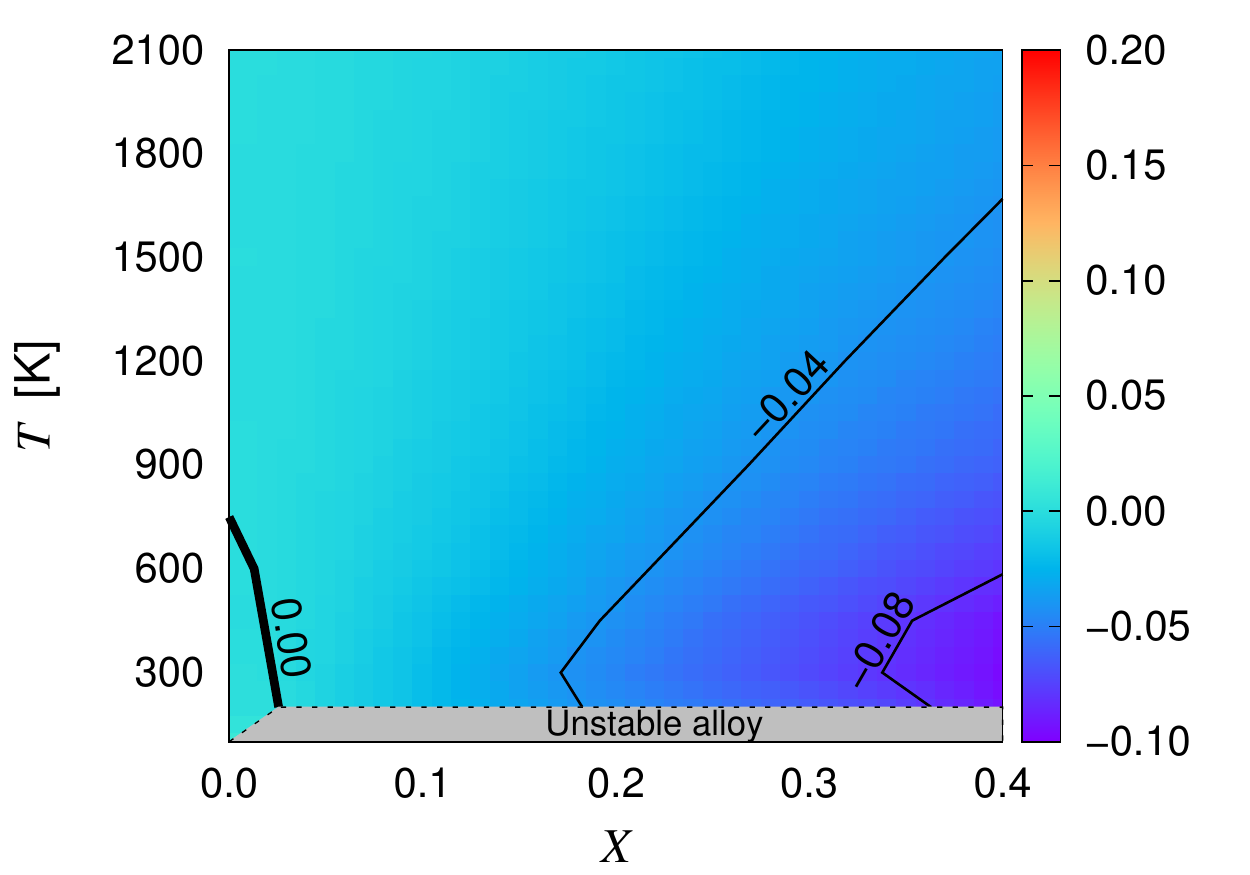}}
	\end{minipage}
	\begin{minipage}{0.32\textwidth}
		\centering
		\subfloat[$C_v=0.1$ at.\%]{\includegraphics[width=0.99\columnwidth]{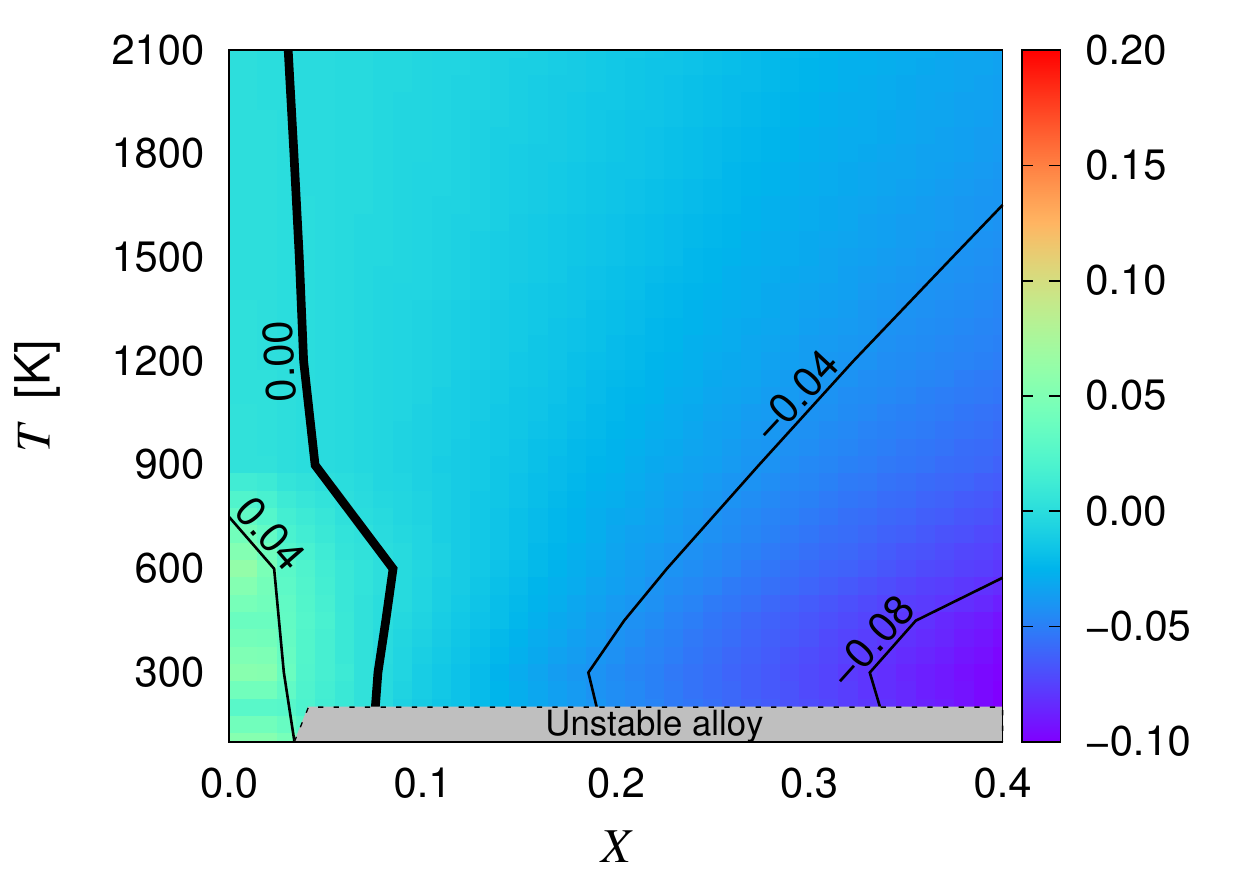}}
	\end{minipage}
	\begin{minipage}{0.32\textwidth}
		\centering
		\subfloat[$C_v=0.2$ at.\%]{\includegraphics[width=0.99\columnwidth]{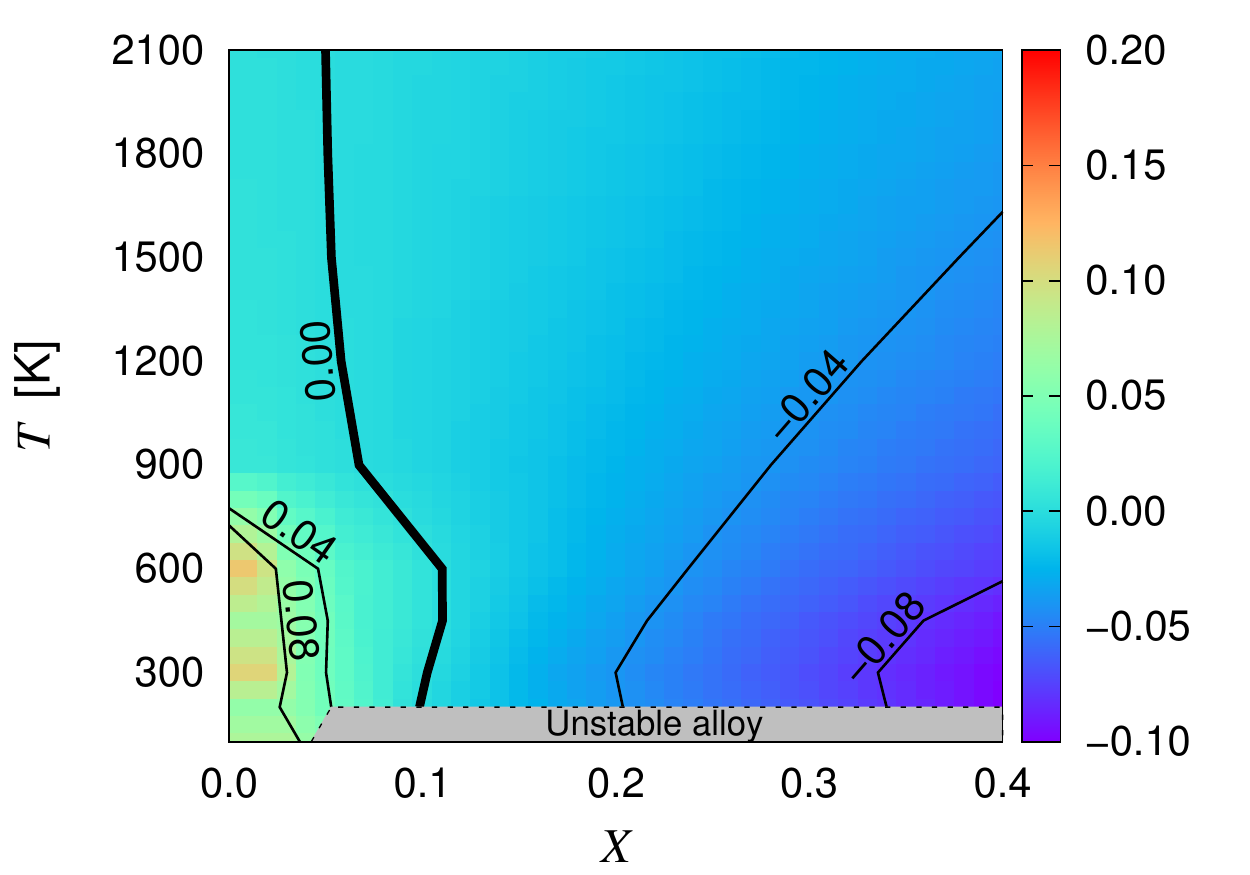}}
	\end{minipage}
	\begin{minipage}{0.32\textwidth}
		\centering
		\subfloat[$C_v=0.5$ at.\%]{\includegraphics[width=0.99\columnwidth]{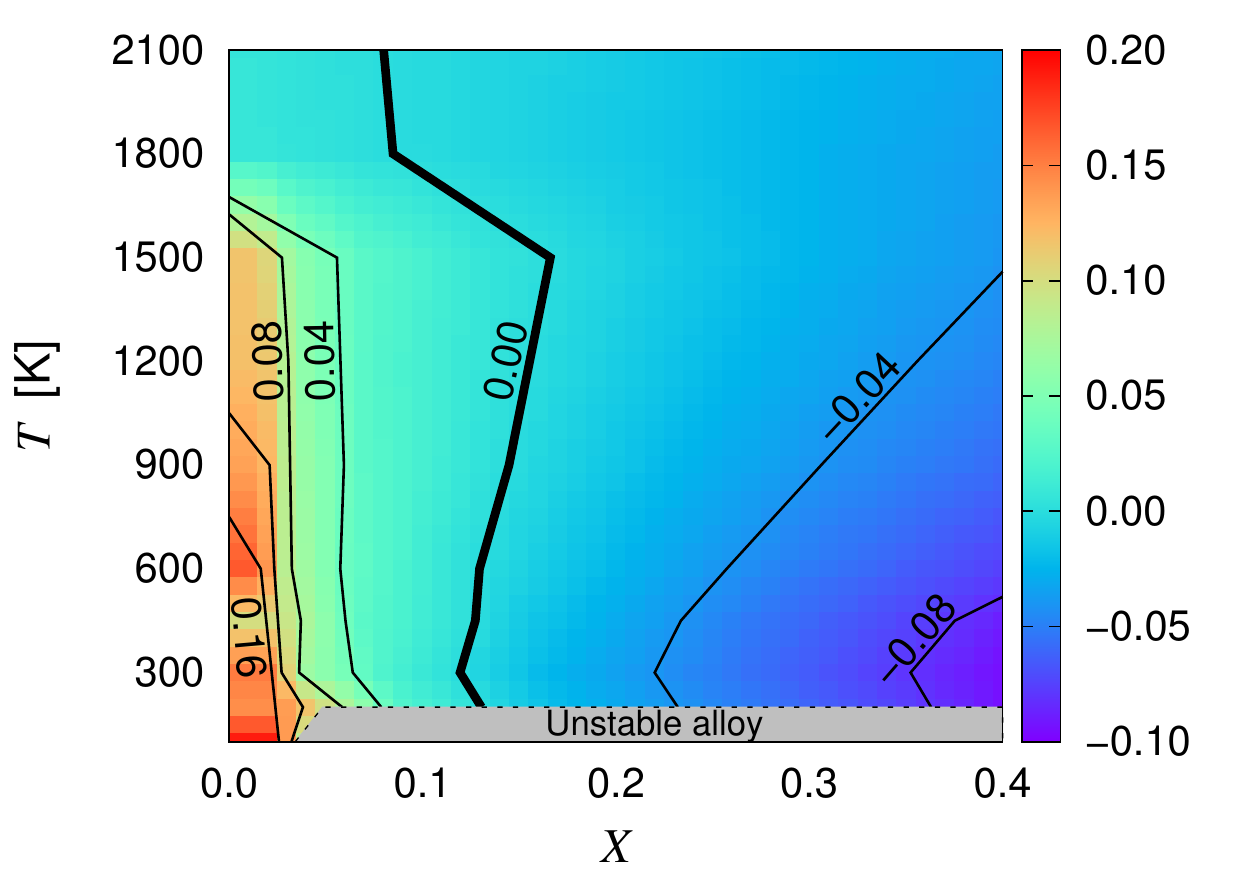}}
	\end{minipage}
\caption{Structural phase diagrams for four different vacancy concentrations. The diagrams clearly show the emergence of regions of solute segregation, characterized by positive SRO and a shifting of the transition phase boundary, $\eta=0$, towards the right (higher concentrations).}
\label{vcc}
\end{figure}

It is well known that non-equilibrium concentrations of defects can alter the thermodynamic behavior of an alloy. For the W-Re system, Wrobel {\it et al.} have studied the ternary W-Re-vacancy system and found that Re clustering occurs in the presence of non-thermodynamic vacancy concentrations \cite{2016arXiv160403746W}. Clusters appear as semi-ordered structures of alternating solute and vacancy planes --a necessity given the short-range repulsion between Re atoms on the one hand, and vacancies on the other (cf.\ Table \ref{dftt}). Next, we carry out a similar study involving various vacancy concentrations, temperatures, and solute concentrations to obtain structural phase diagrams such as that shown in Fig.\ \ref{fig_c000phase}. Each configuration is optimized by combining sgMC steps with energy minimization steps following the process described in Sec.\ \ref{sec_mcABV}. Figure \ref{vcc} shows the diagrams for vacancy concentrations of $C_v=0.01,~0.1,~0.2,~0.5$ at.\% using $64\times64\times64$ primitive cells.

\begin{figure}[h]
\centering
	\begin{minipage}{0.49\columnwidth}
		\centering
		\subfloat[W-1.8at\%Re alloy, 0.5 at\% vacancy concentration.\label{fig_snapV50}]{\includegraphics[width=0.95\textwidth]{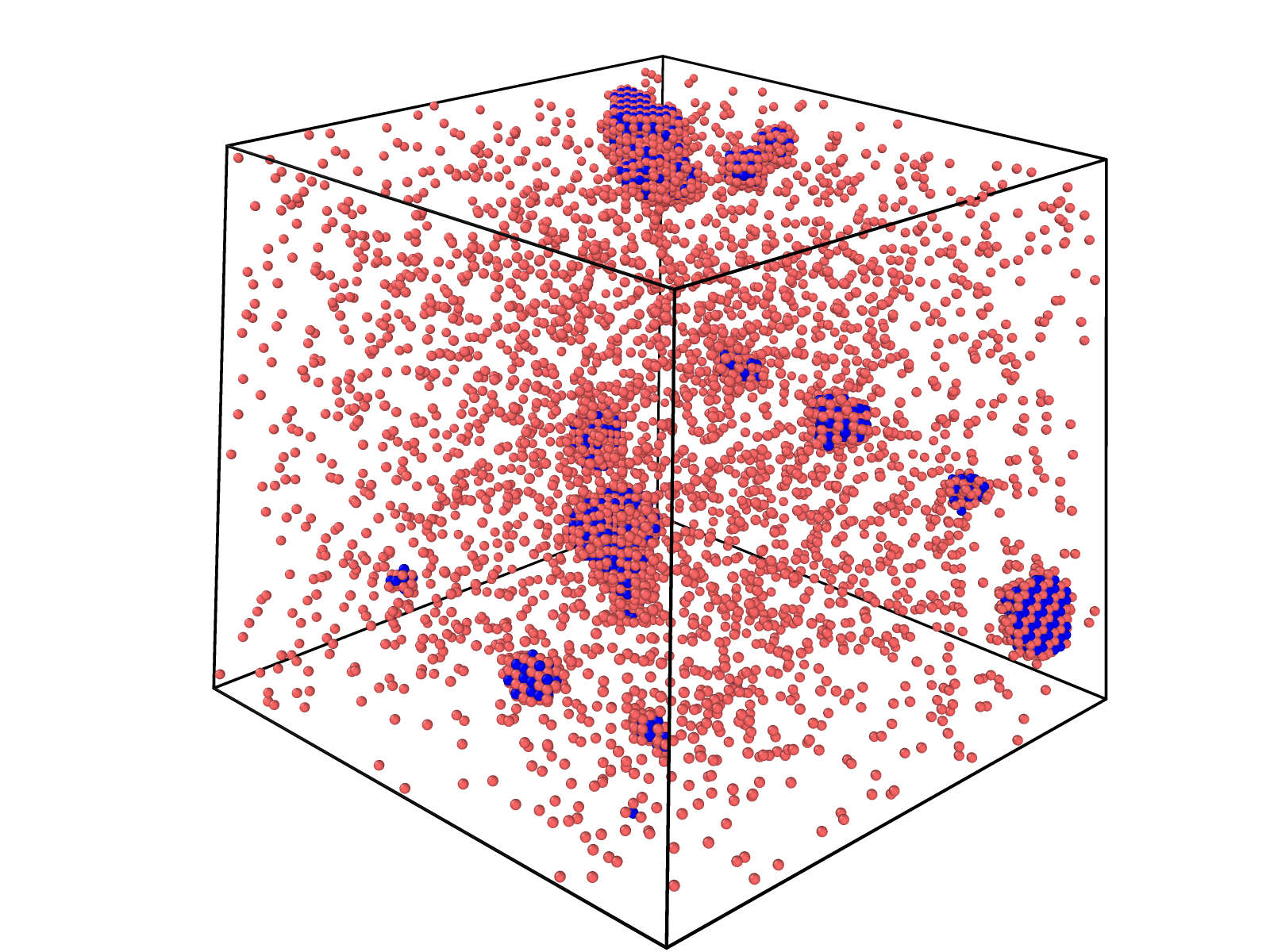}}
	\end{minipage}
	\begin{minipage}{0.49\columnwidth}
		\centering
		\subfloat[W-1.4at\%Re alloy, 0.1 at\% mixed-interstitials.\label{fig_snapI10}]{\includegraphics[width=0.95\textwidth]{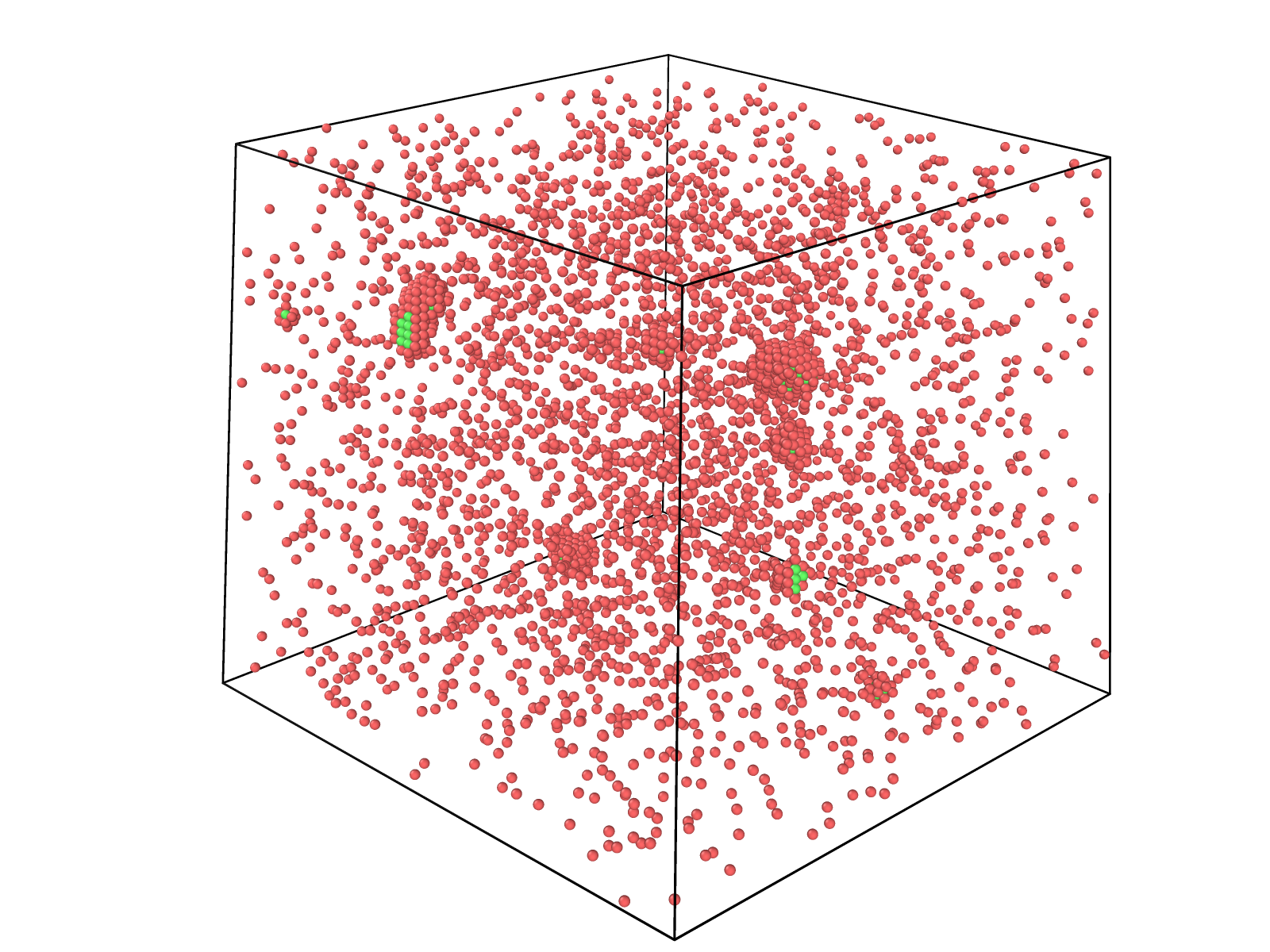}}
	\end{minipage}
	\caption{Equilibrated configurations for W-Re alloys containing different defect concentrations at 600 K. Red spheres represent Re atoms, colored blue or green ones represent the defect in each case.}
\end{figure}

As a representative example, Figure \ref{fig_snapV50} shows the equilibrated configuration at 600 K, 1.8 at.\% Re (which occurs for $\Delta\mu=0.26$), and $C_v=0.5$ at\%.  The figure shows several Re-vacancy clusters with an ordered structure, consistent with the study by Wrobel {\it et al.} \cite{2016arXiv160403746W}. Due to their ordered structure, these solute-vacancy clusters form only at Re concentrations that are commensurate with the vacancy concentration in the system, {\it i.e.}~at values of $X\lesssim0.04$ in most cases.

\subsubsection{Effect of interstitial defects on the phase diagram} \label{sec_sia}
\begin{figure}[tbh]
	\centering
	\includegraphics[width=\columnwidth]{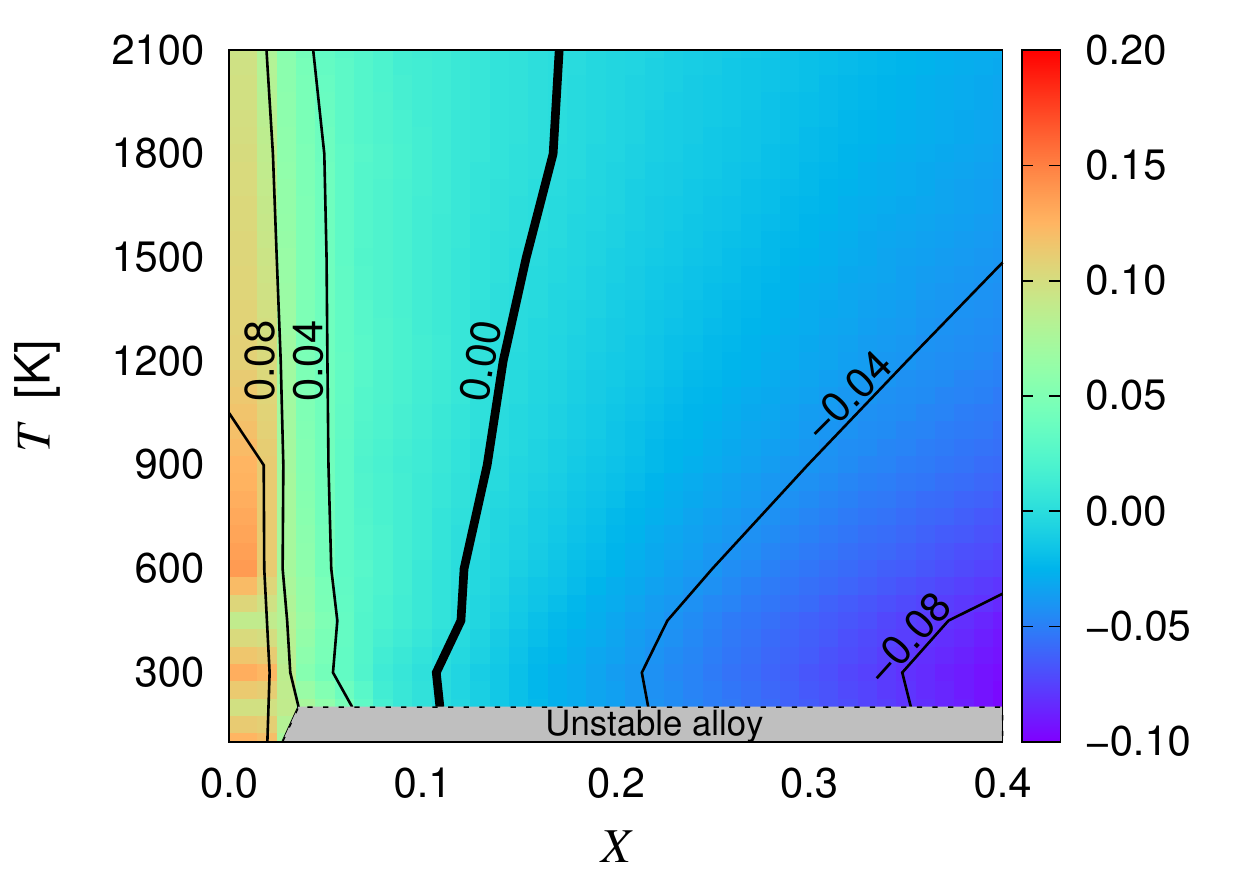}
	\caption{Structural phase diagram for 0.1 at.\% mixed-dumbbell concentration. The diagram shows the emergence of regions of solute segregation, characterized by $\eta>0$, up to $X=0.1\%$.}
	\label{fig_ci10phase}
\end{figure}

Although vacancy concentrations such as those considered in this section are several orders of magnitude larger than the vacancy concentration in thermal equilibrium, one can expect such numbers under far-from-equilibrium conditions such as under high-dose or high-dose rate  irradiation. The case is much more difficult to make for SIAs due to their much higher formation energy (3.2 vs.~10.2 eV, to take two representative numbers \cite{2015JNMGharaee}). However, given the inclination of single interstitials to convert into mixed dumbbells in the presence of solute, it is of interest to repeat the same exercise of looking at the clustering propensity of Re in such cases. The results are shown in Figure \ref{fig_ci10phase} for a defect concentration of 0.1 at.\%. 
The diagram reveals a stronger clustering tendency when interstitials are present compared to vacancies. Such an effect originates from both more attractive binding energies between mixed-interstitials and solute atoms, and between mixed-interstitials with themselves. A snapshot of the equilibrated atomistic configuration is shown in Fig. \eref{fig_snapI10}, where the precipitates are seen to form platelet-like structures with a mixed dumbbell core surrounded by substitutional solute atoms.

\subsection{Kinetic evolution of irradiated W-Re alloys}
There are a number of factors that call for performing kMC simulations in W-Re systems.
\begin{enumerate}
\item First, equilibrium Monte Carlo calculations such as those performed in Section \ref{struct} do not provide information about the precipitate nucleation and growth mechanisms, as well as the timescales involved. 
\item Second, there is clear experimental evidence of Re-cluster formation in the absence of vacancies.  Hasegawa {\it et al}.~\cite{TanHasHe07,HasTanNog11} and Hu {\it et al}.~\cite{Hu2016235} have both reported the formation of W-Re intermetallic precipitates after high-dose, fast neutron irradiation. Moreover, recent irradiation experiments have revealed the formation of Re-rich clusters with bcc structure, {\it i.e.} prior to their conversion into $\sigma$ and/or $\chi$ precipitates. For example, Klimenkov {\it et al}.~note that Re-rich particles not associated with cavities formed in neutron-irradiated single crystal W \cite{Klimenkov2016}. As well, using atom-probe tomography Xu {\it et al}.~have performed detailed analyses of Re-rich atmospheres in bcc W without detecting significant numbers of vacancies \cite{2017ACTAXu}. 
\item New understanding regarding interstitial-mediated solute transport in W-Re alloys \cite{2015JNMSuzudo,2016JAPGharaee}, together with the results in Section \ref{sec_sia}, call for renewed simulation efforts incorporating these new mechanisms --in particular, the three-dimensional and associative nature of Re transport via mixed-dumbbell diffusion.
\end{enumerate}

These considerations motivate the following detailed study of the Re precipitation kinetics under irradiation conditions. 
First, however, we proceed to calculate diffusion coefficients and transport coefficients for defect species and solute atoms.

\subsubsection{Calculation of diffusion coefficients}
\begin{figure}[tbh]
	\centering
		\begin{minipage}{0.45\textwidth}
		\centering
		\subfloat[Vacancy diffusion]{\includegraphics[width=\textwidth]{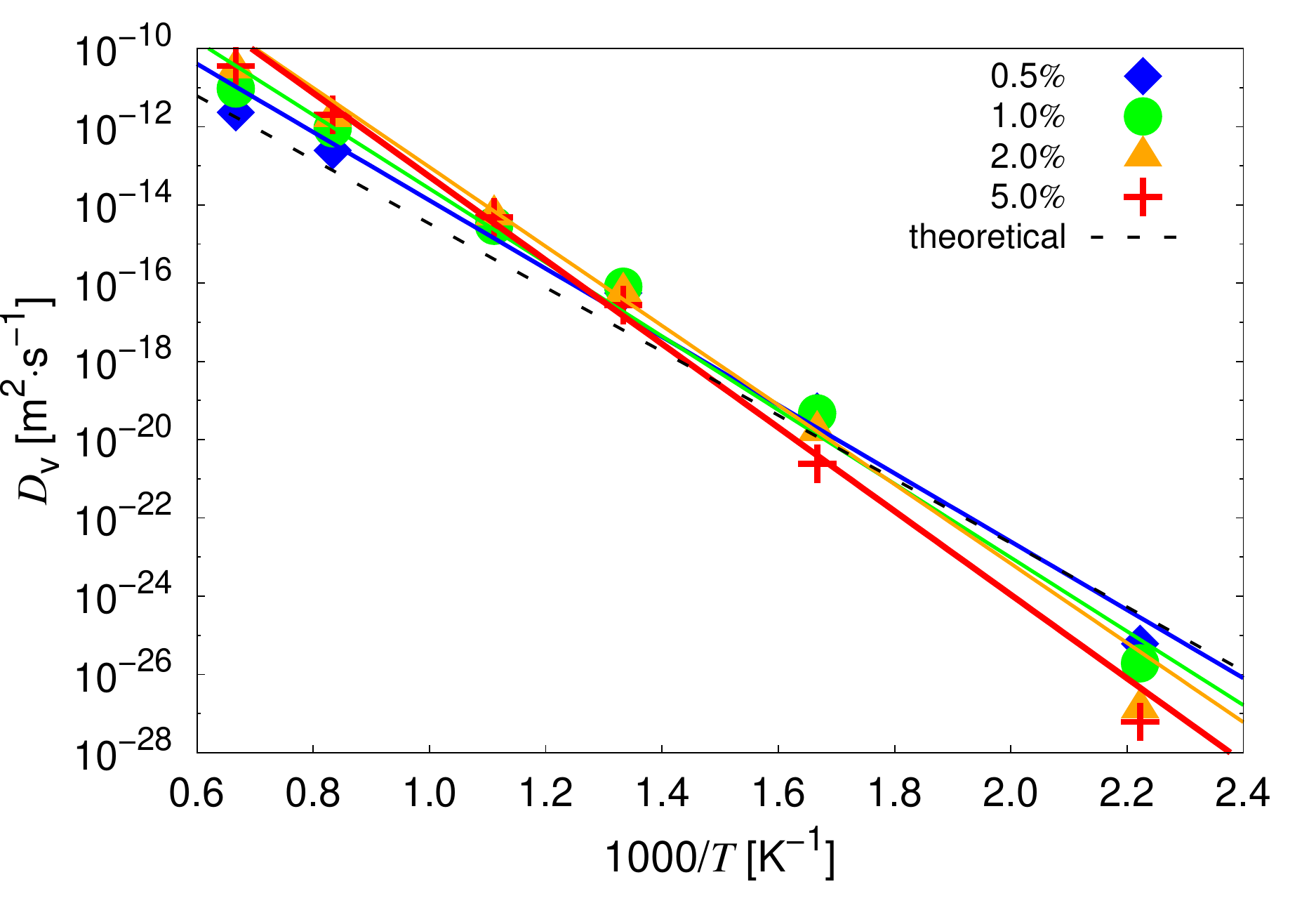}}
	\end{minipage}
	\begin{minipage}{0.45\textwidth}
		\centering
		\subfloat[Solute diffusion]{\includegraphics[width=\textwidth]{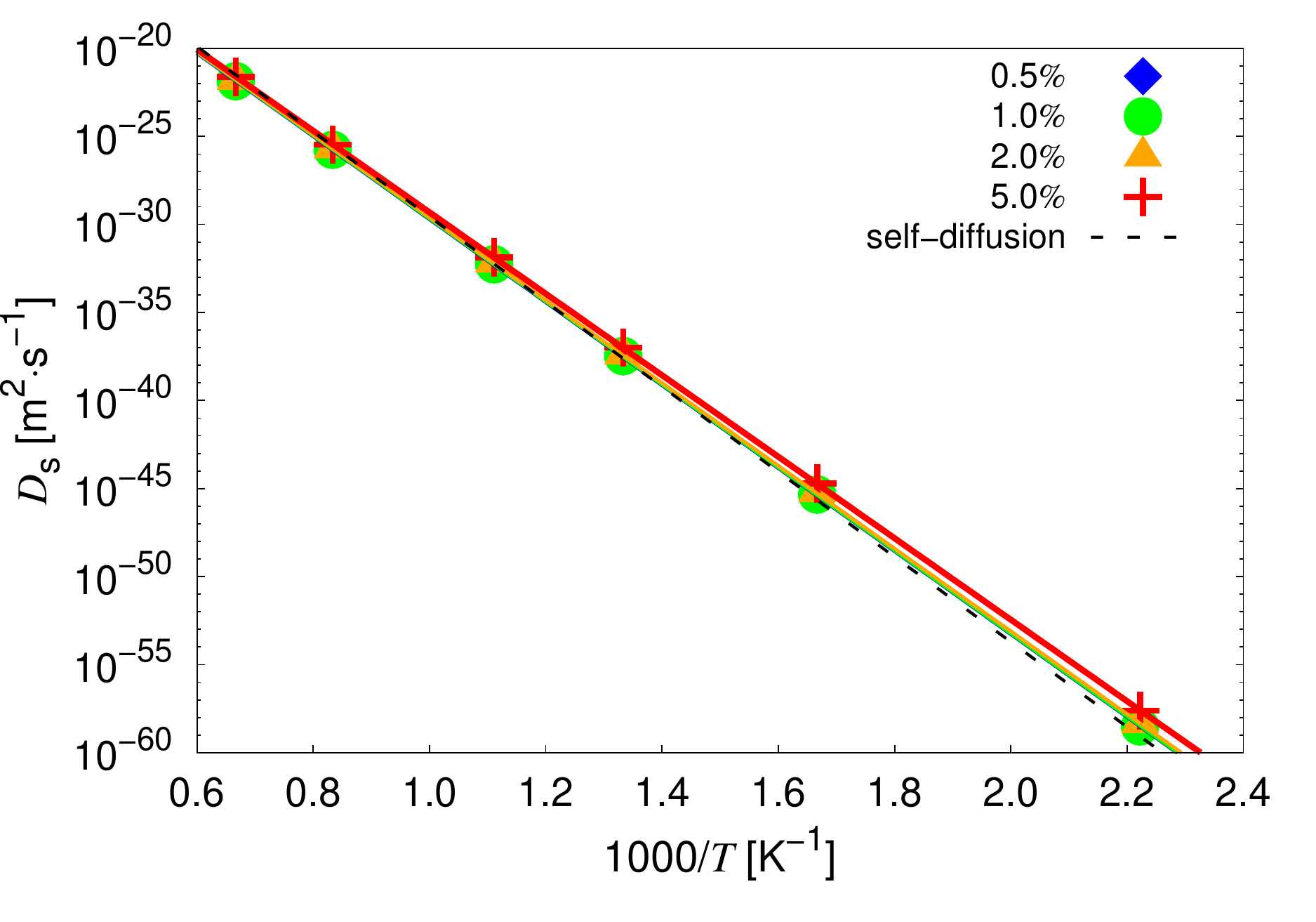}}
	\end{minipage}
	\caption{Diffusivities of vacancies and solute atoms as a function of temperature and alloy concentration. The solid lines correspond to the Arrhenius fits shown in Table \ref{pref}, while the dashed line corresponds to eq.\ \eref{arr}.}
	\label{difffigs}
\end{figure}

Tracer diffusion coefficients (i.e., in the absence of a concentration gradient) for vacancies, interstitials, and solute species in three dimensions are assumed to follow an Arrhenius temperature dependence:
\begin{equation}\label{arr}
D(T)=\nu f\delta^2\exp\left(-\frac{E_a}{kT}\right)
\end{equation}
where $\nu$ is the so-called \emph{attempt} frequency, $f$ is the correlation factor, $\delta$ is the jump distance, $E_a$ is the activation energy, and $D_0=\nu f\delta^2$ is the so-called diffusion pre-factor. Defect diffusivities can be obtained directly from this equation, with $E_a\equiv E_m$. For solute diffusion the above expression must be multiplied times the probability of finding a vacancy in one of the 1nn positions, such that $D_0=z_1\nu f\delta^2$ and $E_a=E_m+E_f^{\rm V}$. However, to calculate the diffusivities of solutes and vacancies as a function of the global solute concentration, fluctuations in local chemistry prevent us from using equations for homogeneous systems such as eq.\ \eref{arr}, and diffusivities must be obtained by recourse to \emph{Einstein}'s equation:
\begin{equation}\label{eins}
	D= \frac{\left<\Delta r^2 \right>}{6\Delta t}
\end{equation}
where $\left< \Delta r^2 \right>$ is the \emph{mean squared displacement} (msd) and $\Delta t$ is the time interval. 
This formula assumes equilibrium defect concentrations, which are generally several orders of magnitude smaller than what a typical simulation cell can afford. For this reason, the time in eq.\ \eref{eins} is not directly the time clocked in the kMC simulations, $\Delta t_{\rm kMC}$. Rather, it must be rescaled by a coefficient that accounts for the difference in defect concentration  \cite{2002_PRBBouar, 2012PRBNastar}:
\begin{equation} \label{eq_timecorr}
    \Delta t= \Delta t_{\rm kMC}\frac{C^{\rm kMC}}{C^{\rm eq}}
\end{equation}
where $C^{\rm kMC}$ and $C^{\rm eq}$ are the defect concentrations in the kMC simulations and in equilibrium, respectively. For simulations involving only one defect, $C^{\rm kMC}$ is simply equal to the inverse of the number of atoms in the computational cell, $C^{\rm kMC}=N^{-1}$, while $C^{\rm eq}= \exp(-E_f/k_BT)$, where $E_f$ is the \emph{instantaneous} defect formation energy, {\it i.e.}~calculated accounting for the local chemical environment. This is the approach used for vacancy mediated diffusion, with $E^{\rm V}_f= \sum_i \epsilon_{\text{V-}\alpha_i}$, where $\alpha_i$ symbolizes the neighboring atoms forming a bond with the vacancy. 
During simulations of solute and vacancy diffusion, $E^{\rm V}_f$ is updated in every Monte Carlo time step and time rescaling is performed on the fly. 
The starting configuration for all calculations involving solute atoms is the equilibrated alloy as obtained in Section \ref{struct} using sgMC simulations.
The results for the vacancy and solute diffusivities, $D_{\rm v}$ and $D_{\rm s}$, can be seen in Figure \ref{difffigs}, while the parameters resulting from fitting the data points in the above figures to eq.\ \eref{arr} are collected in Table \ref{pref}. While $D_{\rm v}$ displays a moderate dependence with the solute concentration, $D_{\rm v}$ is quite insensitive to it. 
\begin{table}[h]
\caption{\label{pref}Diffusion parameters for vacancy and solute diffusion as a function of solute concentration.}
\centering
\begin{tabular}{|c|c|c|}
\hline
$X$ [at. \%] & $D_0$ [m$^2$$\cdot$s$^{-1}$] & $E_m$ [eV] \\
\hline
\multicolumn{3}{|c|}{Vacancy diffusion}\\
\hline
0.0 (eq.\ \eref{arr})    &     $4.84\times10^{-7}$   &   1.62 \\
0.5  &   $6.86\times10^{-6}$  &  1.73 \\
1.0  &   $6.92\times10^{-5}$  &  1.87 \\
2.0  &   $1.26\times10^{-3}$  &  2.08 \\
5.0  &   $2.57\times10^{-3}$  &  2.16 \\
\hline
\multicolumn{3}{|c|}{Solute diffusion}\\
\hline
0.0 (eq.\ \eref{arr})   &    $3.87\times10^{-6}$  & $1.62+3.17 = 4.79$ \\
0.5 &   $7.56\times10^{-7}$  &  4.67 \\
1.0 &   $7.80\times10^{-7}$  &  4.67 \\
2.0 &   $7.89\times10^{-7}$  &  4.66 \\
5.0 &   $6.75\times10^{-7}$  &  4.59 \\
\hline
\end{tabular}
\end{table}

As discussed in Sec.\ \ref{sec_mig}, self-interstitial migration occurs by way of fast sequences of $\langle111\rangle$ transitions punctuated by sporadic rotations, whereas mixed dumbbell diffusion occurs via random $\langle100\rangle$ hops in three dimensions. 
Interstitial diffusivities of both types can be calculated straightforwardly by using eq.\ \eref{arr} parameterized with the data in Table \ref{tab_mig}.

\subsubsection{Calculation of transport coefficients}\label{sec:trans}

\begin{figure}[tbh]
	\centering
		\begin{minipage}{0.45\textwidth}
		\centering
		\subfloat[Solute-solute transport coefficient\label{lss}]{\includegraphics[width=\textwidth]{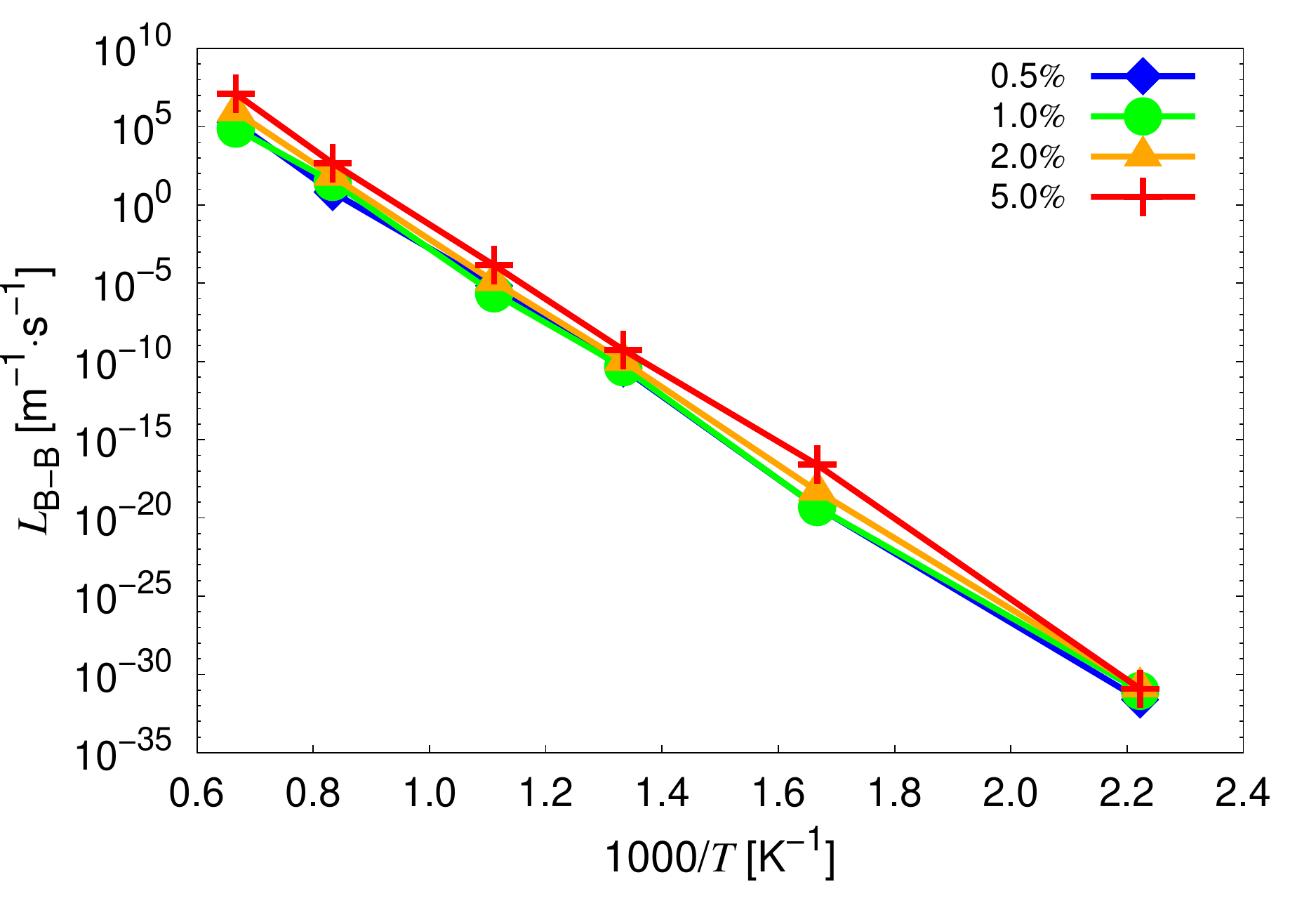}}
	\end{minipage}
	\begin{minipage}{0.45\textwidth}
		\centering
		\subfloat[Solute-vacancy transport coefficient\label{lsv}]{\includegraphics[width=\textwidth]{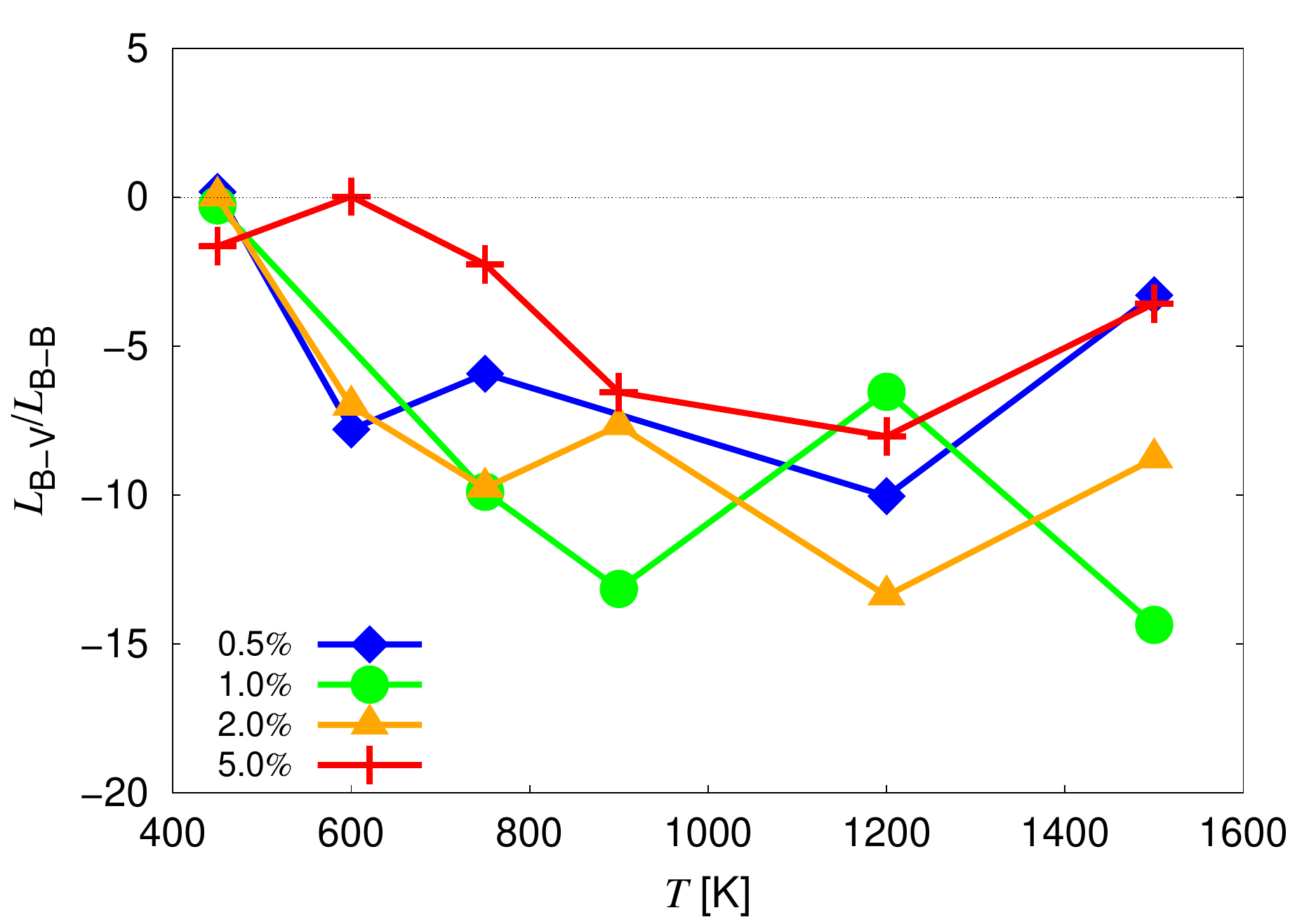}}
	\end{minipage}
	\caption{Phenomenological transport coefficients for solute-solute and vacancy-solute interactions.}
\end{figure}

Within linear response theory, mass transport can be related to chemical potential gradients via Onsager's phenomenological coefficients. The value and sign of these transport coefficients can provide important physical information about the nature of solute and defect fluxes. On a discrete lattice, the transport coefficients $L_{ij}$ coupling two diffusing species can be calculated as \cite{2016ActaSenninger, 2003BOOKallnatt}:
\begin{equation}
	L_{ij}= \frac{1}{6V} \frac{\left< \Delta r_i \Delta r_j \right>}{\Delta t}
\end{equation}
where $V$ is the total volume of the system; $\Delta r_i$ is the total displacement of species $i$, and $\Delta t$ is the rescaled time. Here we focus on the relationship between solutes and vacancy and solute atoms, $L_{\rm B\text{-}B}$, and $L_{\rm B\text{-}V}$, as a function of temperature and solute content. Due to the associative transport mechanism of AB interstitials, the corresponding transport coefficient relating interstitials with solute atoms is always positive and we obviate its calculation. 
Figure \ref{lss} shows the results for $L_{\rm B\text{-}B}$, which displays an Arrhenius temperature dependence and is always positive. The dependence with solute concentration is not significant up to 5\%, with an average activation energy of 4.7 eV --very similar to the solute diffusion activation energy-- and a prefactor of approximately $3.9\times10^{20}$ m$^{-1}$$\cdot$s$^{-1}$. $L_{\rm B\text{-}B}$ is by definition related to the solute diffusion coefficient presented above.

In Figure \ref{lsv} we plot the ratio $L_{\rm B\text{-}V}/L_{\rm B\text{-}B}$. Two observations stand out directly from the figure. First, the value of $L_{\rm B\text{-}V}$ is always negative (the exception being at 450 K, when is almost zero). This indicates a reverse coupling between solutes and vacancies, {\it i.e.}~vacancy fluxes would oppose solute fluxes. The implications of this calculation will become clearer when we study solute precipitation in the next section. Second, $L_{\rm B\text{-}V}$ is on average about an order of magnitude larger than $L_{\rm B\text{-}B}$,which is to be expected for substitutional solutes moving by a vacancy mechanism.

\subsubsection{Kinetic Monte Carlo simulations}
To narrow down the large parametric space associated with irradiation of W-Re alloys (Re concentration, temperature, dose, dose rate, etc.), we resort to the study performed in Secs.\ \ref{struct} and \ref{sec_defect}. It was seen there that segregation occurs most favorably at low solute compositions. For this reason, and to enable comparison with the work by Xu {\it et al.} \cite{XuBecArm15,2017ACTAXu}, we choose a W-2.0\% at.~Re alloy for our study. By way of reference, this would correspond to the transmutation fraction attained after a dose of 12 dpa or 4 full-power years in DEMO's first wall according to Gilbert and Sublet \cite{GilSub11}. 
When relatively high concentrations of defects are present --as one might expect during irradiation-- precipitation is also favored at high temperatures, so here we carry out our simulations between 1700 and 2000 K. 
We use box sizes of $64^3$ and $80^3$ with a damage insertion rate of $10^{-3}$ dpa per second. As shown in Appendix \ref{appendix}, the equivalence relation that exists between both box sizes  enables us to compare them directly. The three different defect-sinks discussed in Sec.~\ref{sec_mcABVI} are all considered here.

\begin{figure}[tbh]
	\centering
	\includegraphics[width=\columnwidth]{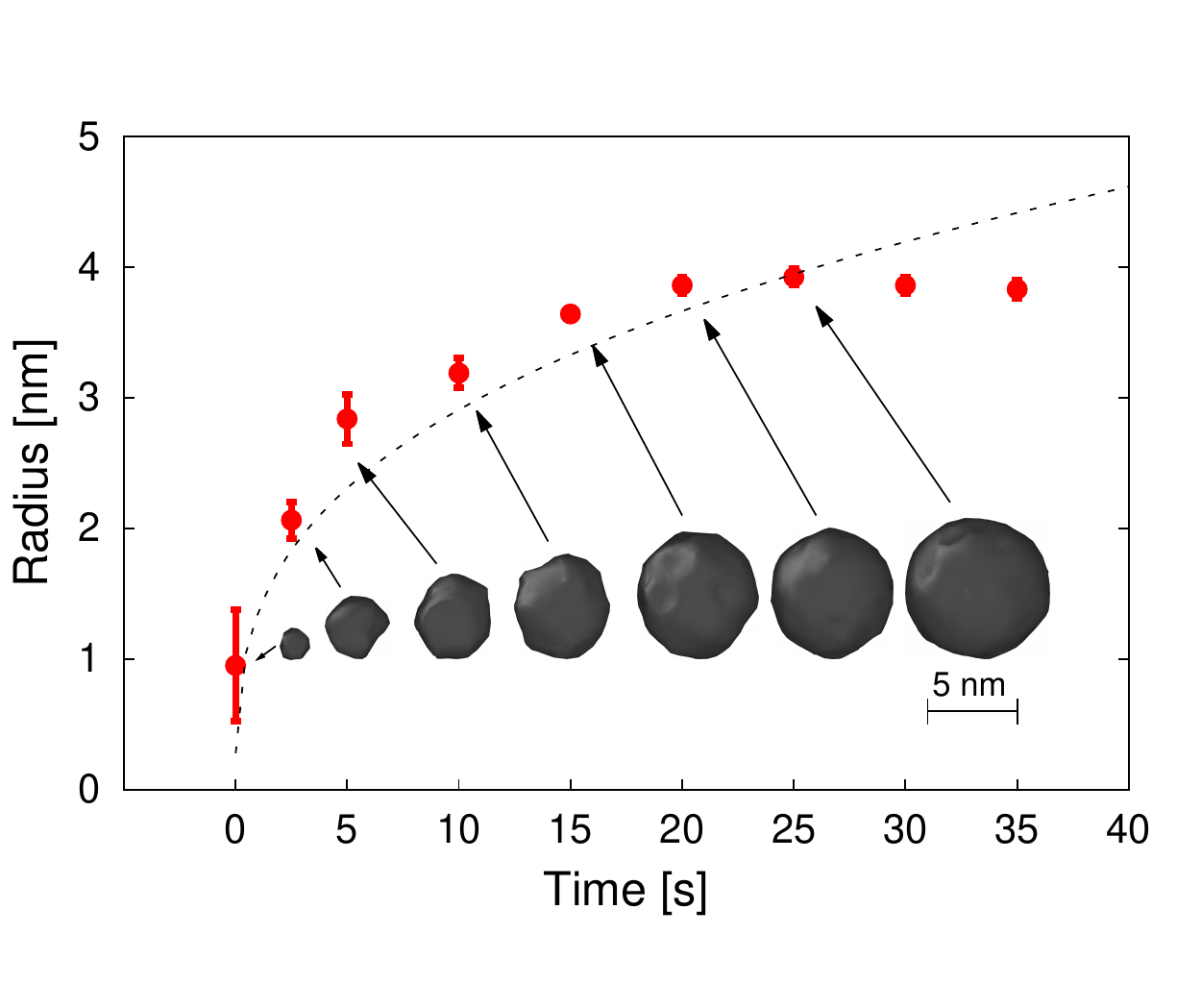}
	\caption{Precipitate growth with time at 1800 K and $10^{-3}$ dpa$\cdot$s$^{-1}$ in a W-2.0\% at.~Re alloy. The dashed line represents perfect spherical growth (cf.\ \ref{appendix}). A surface reconstruction rendition of one precipitate at various times is provided as inset.}
	\label{radius}
\end{figure}

We first investigate the kinetic evolution of a system with no sinks. Eight independent simulations were conducted.
It is seen that after an average waiting time of $\approx21$ seconds (or $\approx0.02$ dpa) one precipitate starts to grow in all cases. This time can be regarded as the average \emph{incubation} (nucleation) time for the conditions considered in the study. Figure \ref{radius} shows the mean size from all eight cases as a function of \emph{growth} time, {\it i.e.}~initializing the clock after the cluster nuclei are formed regardless of the observed incubation time. The dashed line in the figure is the associated spherical growth trend, which the precipitates are seen to follow for approximately 20 s. Subsequently, growth stops at a saturation radius of 4 nm, which is seen to be the stable precipitate size.
A surface reconstruction rendition of one of the precipitates is also provided in the figure as a function of time. This depiction as a compact convex shape is not intended to represent the true diffuse nature of the cluster, and is only shown as an indication of the cluster average size and shape.

\begin{figure}[tbh]
	\centering
	\includegraphics[width=\columnwidth]{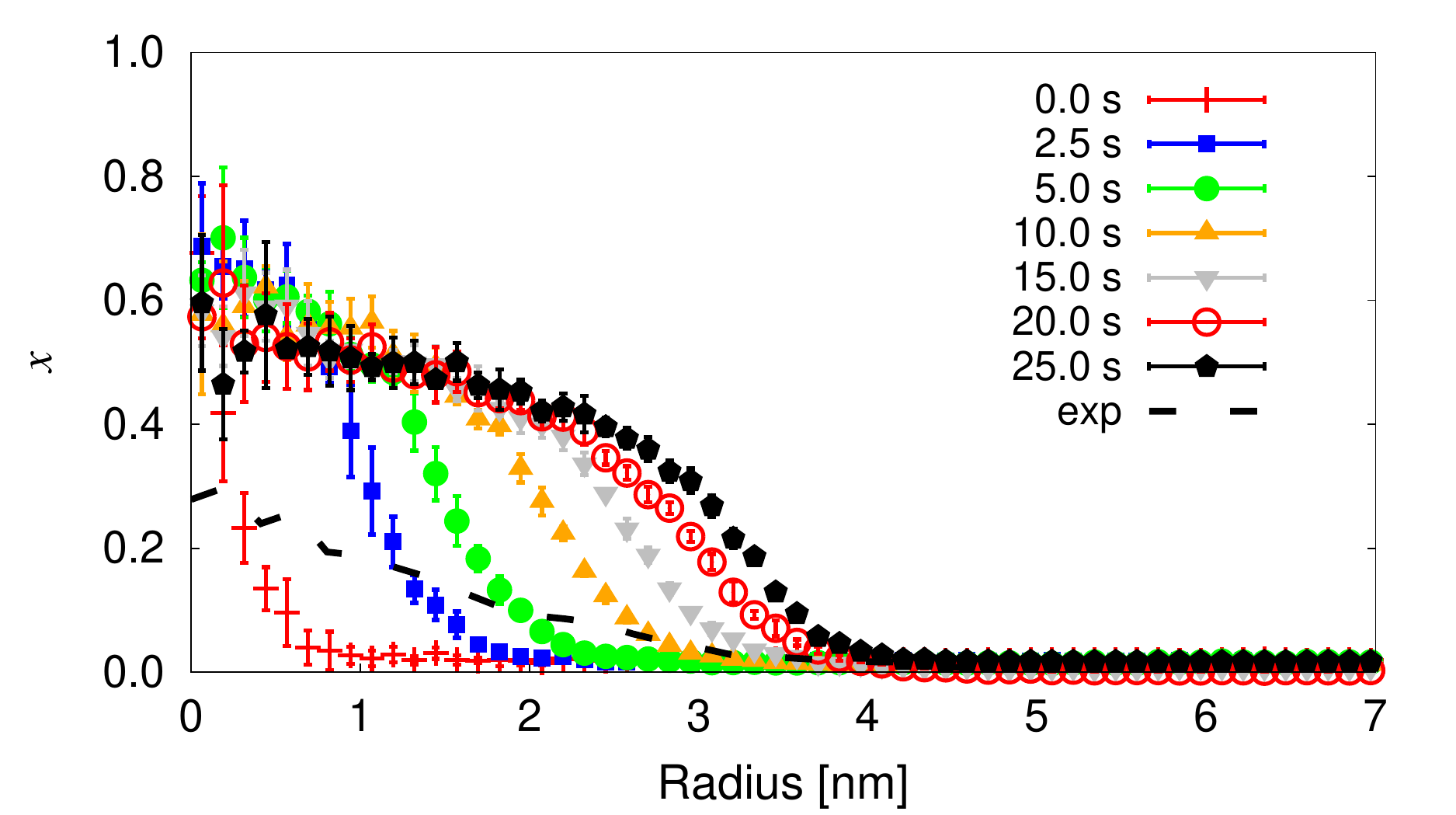}
	\caption{Radial concentration profile as a function of time for the precipitates formed in the kMC simulations. The experimental results are taken from the work by Xu {\it et al.}~\cite{2017ACTAXu}.}
	\label{proff}
\end{figure}

The next question we address is the solute concentration inside the precipitate. Xu {\it et al.}~\cite{XuBecArm15,2017ACTAXu} have performed detailed atom probe analyses of radial concentration profiles at 573 and 773 K and find that the precipitates that form might be better characerized as 'solute clouds', reaching concentrations of around 30\% in the center gradually declining as the radius increases. Our analysis is shown in Figure \ref{proff}, with results averaged over the 8 cases tried here. The figure shows that the concentration at the precipitate core (within the inner 1.5 nanometers) surpasses 50\% --the thermodynamic limit for the formation of intermetallic phases--, which could provide the driving force for such a transformation. Because our energy model is not valid above the solid solution regime, we limit the interpretation of such phenomenon however. What is clear is that the precipitates are not fully-dense, even near their center. In fact, the relative solute concentration appears to diminish near the precipitate core once the saturation point has been reached.

\begin{figure}[tbh]
	\centering
	\includegraphics[width=\columnwidth]{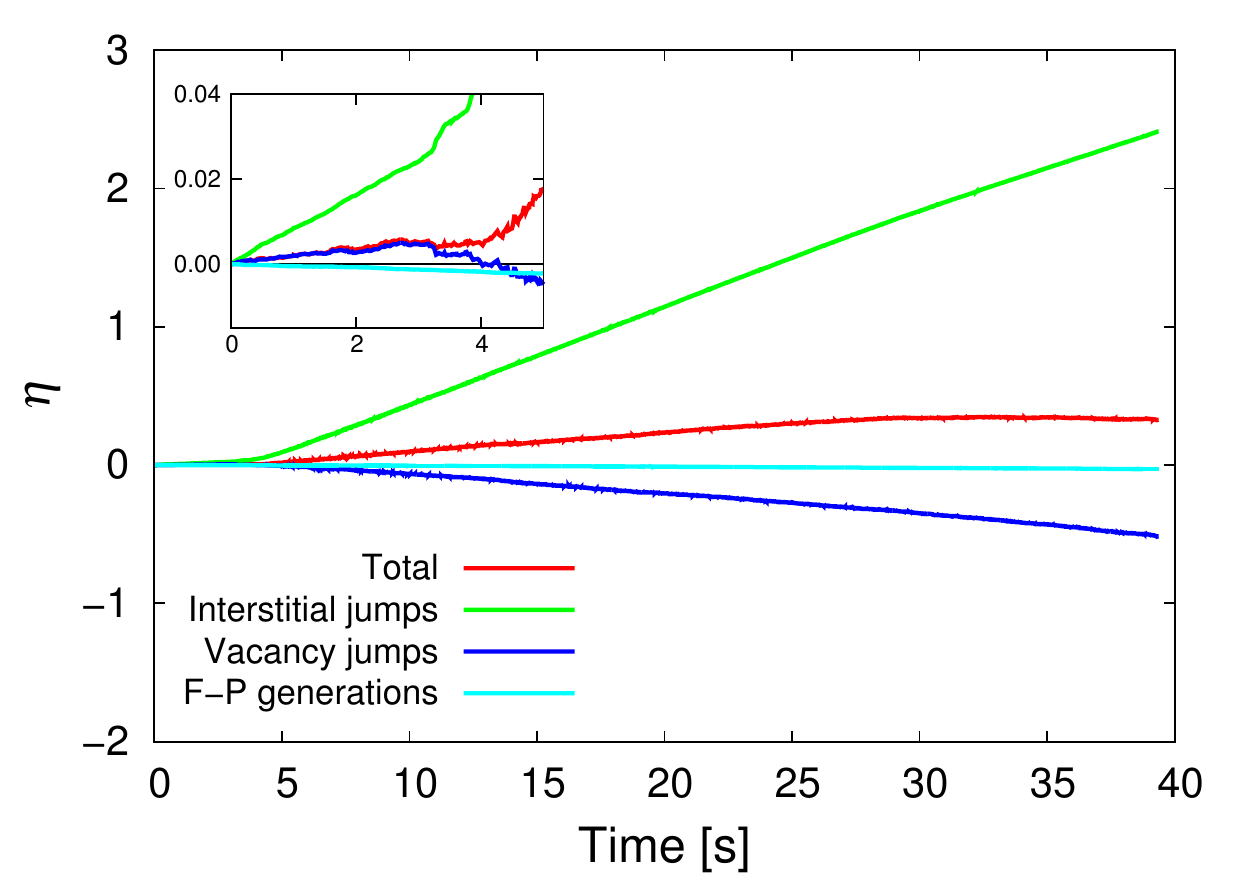}
	\caption{Evolution of the differential SRO during the nucleation and growth in the kMC simulations.}
	\label{srotime}
\end{figure}

Finally, we address the issue of whether it is vacancy or interstitial mediated transport that is primarily responsible for solute agglomeration and the formation of Re-rich clusters. To this end, we track the evolution with time of the incremental SRO change brought about by any given kMC event during the formation of one the precipitates discussed above.
The results are given in Figure \ref{srotime}, where contributions from SIA and mixed interstitial jumps, vacancy jumps, and Frenkel pair insertion are plotted. 
These results conclusively demonstrate that mixed-interstitial transport is dominant among all other events to bring solute together. 
Vacancies, on the other hand, serve a dual purpose. They first act as a `hinge' between solute atoms that would otherwise repel, much in the manner shown in Fig.\ \ref{fig_snapV50}. This results in an initial positive contribution to the SRO, as shown in the inset to Fig.\ \ref{srotime}, by forming dimers, trimers, or other small solute clusters. However, once a critical nucleus forms and starts to grow, vacancies reverse this behavior and act to dissolve the precipitate (differential SRO turns negative in Fig.\ \ref{srotime}), mostly by making the precipitate/matrix interface more diffuse.
As expected, Frenkel pair insertion has practically no effect on the overall precipitate evolution. 

\begin{figure}[tbh]
	\centering
	\begin{minipage}{0.32\textwidth}
		\centering
		\subfloat[During cluster nucleation.\label{rec1}]{\includegraphics[width=\textwidth]{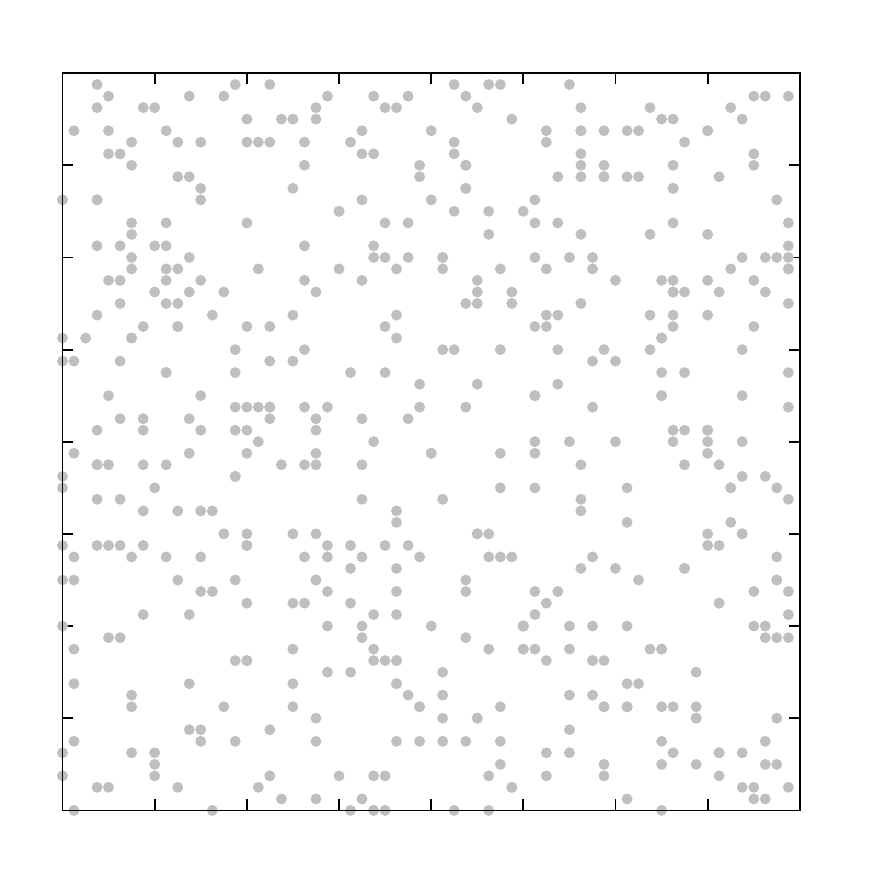}}
	\end{minipage}
	\begin{minipage}{0.32\textwidth}
		\centering
		\subfloat[During precipitate growth\label{rec2}]{\includegraphics[width=\textwidth]{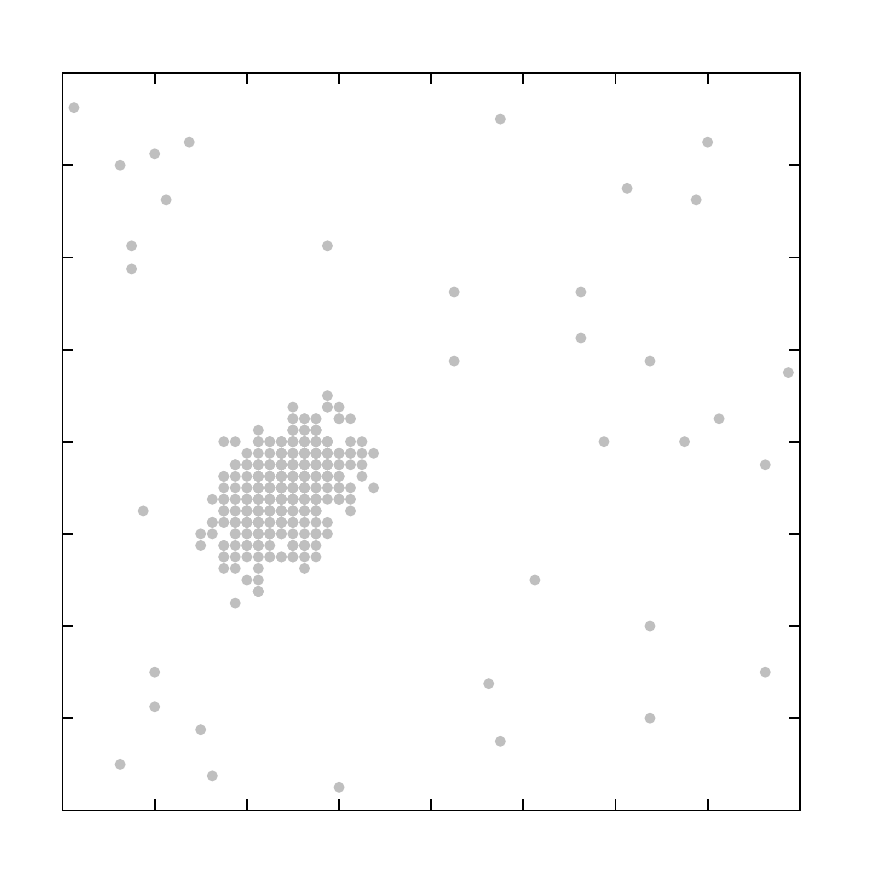}}
	\end{minipage}
	\begin{minipage}{0.32\textwidth}
		\centering
		\subfloat[After size saturation.\label{rec3}]{\includegraphics[width=\textwidth]{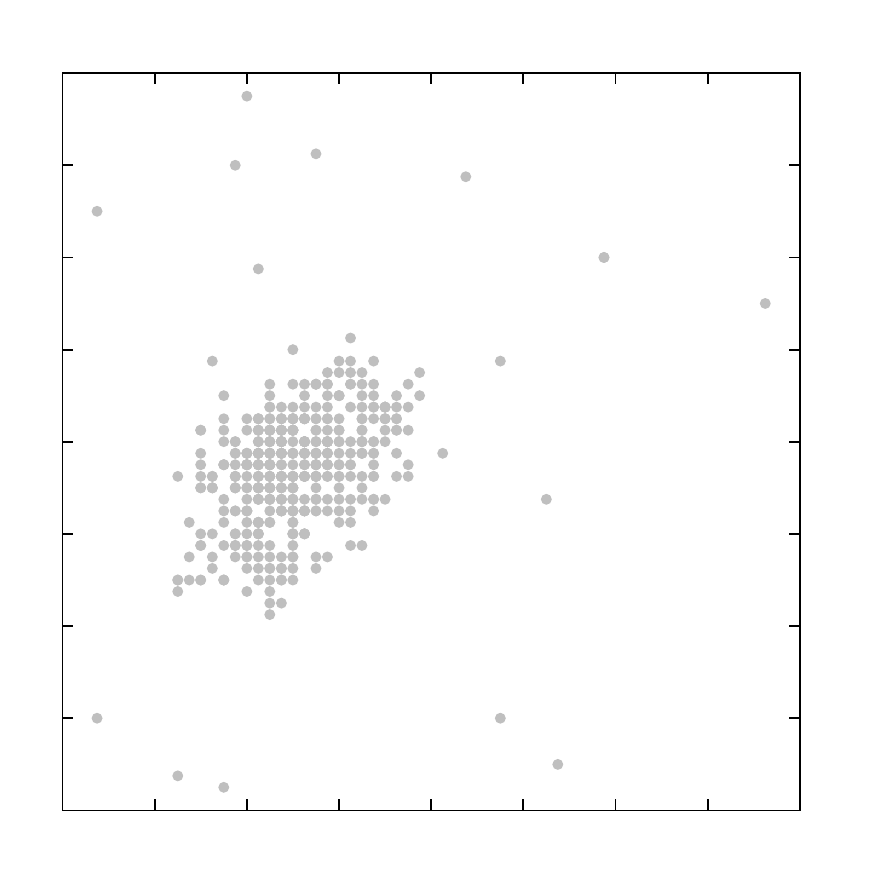}}
	\end{minipage}
	\caption{Spatial distribution of recombination events for several stages of precipitate evolution.}
	\label{rec}
\end{figure}

The precipitate grows by a sustained capture of mixed interstitials and subsequent attraction of vacancies. This gives rise to localized recombination at the precipitate, which makes the precipitates incorporate solute atoms over time. Figure \ref{rec} shows the spatial location of the recombination events during a period of 2.0 s before, during, and after precipitate growth. The figure clearly shows that, once formed, the precipitate becomes a preferential site for recombinations, which results in further growth and eventually in saturation. Because the primary source of solute is via interstitial transport, which also brings W atoms, the precipitates are never fully compact ($x\sim1$). Instead, maximum concentrations of around 50\% are seen near the center when the precipitates reach their saturation size of 4-nm radius. As we will discuss in the next section, this is consistent with experimental measurements and observations of both coherent bcc clusters and incoherent $\sigma$ and $\chi$ phases.


Simulations performed with defect sinks in the same temperature range simply result in solute segregation in the manner described in our previous work \cite{2016JPCMHuang}. Radiation induced precipitation results from the onset of defect fluxes to the sinks, providing sufficient competition to delay the formation of bulk precipitates beyond the time scales coverable in our kMC simulations. More information is provided in the Supplementary Information.

\section{Discussion and conclusions} \label{sec:disc}
\subsection{Mechanism of nucleation and growth}

On the basis of our results, the sequence of events that leads to the formation of Re-rich precipitates in irradiated W-2Re (at\%) alloys is as follows: 

\begin{enumerate}
\item First, a Frenkel pair is inserted in the computational box following the procedure specified in Sec.\ \ref{sec_mcABVI}.
As interstitials enter the system, they perform a one-dimensional migration until they encounter a solute atom, after which they become mixed AB dumbbells capable of transporting solute in 3D. As these mixed dumbbells diffuse throughout the lattice, they encounter substitutional solute atoms and become trapped forming a B-AB complex with a binding energy of 0.15 eV (cf.\ Table \ref{dftt}).

\item The vacancy in the Frenkel pair migrates throughout the lattice contributing to the formation of small Re complexes (dimers, trimers, tetramers, etc.). Vacancy motion does not necessary imply solute drag, as indicated by the negative value of transport coefficients in Sec.\ \ref{sec:trans}. However, as the evidence from the Metropolis Monte Carlo simulations in Fig.\ \ref{vcc} suggests, they can form small complexes of stable Re-V clusters.

\item The vacancy can become locally trapped in the small Re-V complexes mentioned above. However, at the high temperatures considered here, it is likely to de-trap and continue migrating until it finds the immobilized interstitial from (1), as this provides the largest thermodynamic driving force to reduce the energy of the crystal. When the vacancy and the interstitial meet, another small V-Re cluster is formed. 
Throughout this process, both mixed interstitial and vacancy hops are characterized by an increasing differential SRO parameter (cf.\ Fig.\ \ref{srotime}).

\item Eventually, one of these Re clusters grows larger than the rest due to natural fluctuations. When that happens the likelihood that the V-AB recombination will take place at that larger cluster grows. This signals the onset of the growth process, fueled by continued attraction of AB mixed dumbbells and the subsequent associated recombination. At this stage, vacancies reverse their role as solute-atom `hinges' and begin to contribute to cluster dissolution (negative differential SRO parameter in Fig.\ \ref{srotime}). This results in the development of a more or less diffuse interface as the precipitate grows, which delays the next recombination event and slows down growth.

\item Although the precipitate continues to be the main pole of attraction for vacancy-interstitial recombinations (cf.\ Fig.\ \ref{rec}), the system reaches a point where most of the solute is consumed into a diffuse precipitate that halts further growth. Vacancies then have more time to interact with the interface atoms before the next recombination event, which results in a smearing of the precipitate interface. In the absence of sinks, or other precipitates, the existing cluster is the sole focus of solute agglomeration, which allows it grow to its maximum size for the current alloy content of 2\% Re. It is to be expected that with competing solute sinks, the precipitates might either be slightly smaller in size or less solute-dense internally.

\end{enumerate}
This qualitative explanation is built on direct evidence and interpretation from our results, described in detail in Sec.\ \ref{sec:res}. However, to support some of the above points more explicitly, we provide additional details as Supplementary Information.

Interestingly, the essential features of our mechanism were originally proposed by Herschitz and Seidman \cite{herschitz1984atomic,herschitz1985radiation} on the basis of atom probe observations of neutron-irradiated W-25Re alloys. Remarkably, these authors had the intuition to propose the basic elements needed to lead to Re precipitate formation identified in our work with the more limited understanding available at the time.

\subsection{Brief discussion on the validity of our results}

Our simulations are based on a highly-optimized implementation of the standard kMC algorithm. With the computational resources available to us, we can reasonably simulate systems with less than 500,000 atoms into timescales of tens of seconds. This has proven sufficient to study Re clustering at high temperatures, where vacancy mobility is high and comparable to mixed-interstitial mobility. Recall from the previous section that the formation of clusters is predicated on the concerted action of both defect species, with mixed interstitials becoming trapped at small Re clusters followed by a recombination with a vacancy that makes the cluster grow over time. Clustering and precipitation of Re in irradiated W has been seen at temperatures sensibly lower than those explored here, such as 573 and 773 K for ion-irradiated W-Re \cite{XuBecArm15,Edmondson15,2017ACTAXu}, 773 and 1073 K for neutron irradiated W in HFIR \cite{Hu2016235}, 1173 K in neutron irradiated W in the HFR reactor \cite{Klimenkov2016}, and Williams {et al.} at 973$\sim$1173 K in EBR-II \cite{WilWifBen83}. The work by Hasegawa {\it et al.} in JOYO \cite{TanHasFuj08,HasTanNog11} does cover --by contrast-- a similar temperature range as ours. The principle is that the mechanism proposed here can be conceivably extended to lower temperatures without changes with just a timescale adjustment due to the significantly slower mobility of vacancies at those temperatures. This would require simulated times that are far too long to cover with kMC.

An intrinsic limitation of our model is that it is based on a rigid bcc lattice and cannot thus capture the transition of precipitates to the intermetallic phase. As such, our model does not necessarily reflect the true microstructural state when the local concentration surpasses 40$\sim$50\%, which is when phase coexistence is expected to occur according to the phase diagram \cite{2000JoNMEkman}. However, our simulations are useful to determine the kinetic pathway towards the accumulation of Re concentrations in the vicinity of that amount. Neutron irradiation experiments such as those performed at JOYO and HFIR reveal the formation of acicular $\sigma$ and $\chi$ precipitates \cite{WilWifBen83,TanHasFuj08,HasTanNog11,Hu2016235}, which presumably indicates reaching local values of Re concentration of or higher than 40$\sim$50\% at the site of precipitate formation. However, in controlled ion irradiation experiments \cite{XuBecArm15,Edmondson15,2017ACTAXu} there is clear evidence that the precursor to the formation of these intermetallic precipitates are noncompact Re-rich clusters with bcc structure. We cannot but speculate how the transition from these solute-rich clusters to well-defined line compounds $\sigma$ and $\chi$ takes place (perhaps via a martensitic transformation, as in Fe-Cu systems \cite{ErhMarSad13}), but it is clear that it is preceded by the nucleation and growth of coherent Re clusters. In our simulations, we find that the clusters have a maximum concentration of $\approx$50\% in the center, in contrast with Xu {\it et al}., who observe concentrations no larger than 30\%. This disparity may simply be a consequence of the different temperatures considered relative to our simulations (773 vs 1800 K), as it is expected that the accumulation of solute by the mechanism proposed here will be accelerated by temperature.

As well, our Re clustering mechanism is predicated on the insertion of Frenkel pairs, when it is well known that fast neutron and heavy-ion irradiation generally result in the formation of clusters of vacancies and interstitials directly in displacement cascades. However, even here tungsten is somewhat of a special case. Recent work \cite{Sand2013,Yi2015,Setyawan2015329} suggests that most of the defects in high-energy ($>$150 keV) cascades in W appear in the form of isolated vacancies and interstitials. This, together with the fact that most displacement cascades for non-fusion neutrons and heavy ions have energies well below the 150-keV baseline, gives us confidence that our mechanism would be operative even in such scenarios.

\subsection{Implications of our study}
Beyond the obvious interest behind understanding the kinetics of Re-cluster formation in irradiated W-Re alloys, our model is useful to interpret other physical phenomena. For example, it is well known that swelling is suppressed in irradiated W-Re alloys compared to pure W \cite{Matolich1974837}. By providing enhanced avenues for interstitial-vacancy recombination, small Re clusters capture mixed interstitials, allowing sufficient time for vacancies to subsequently find them and suppressing the onset of swelling. Intrinsic 3D mobility of mixed dumbbells is likely to favor recombination as well. Note that this explanation for swelling suppression is different to the one proposed for Fe-Cr alloys, where 1D migration of SIAs is restrained by Cr atoms \cite{terentyev2005correlation}.

Finally, the mechanisms proposed here refer to homogeneous nucleation, {\it i.e.}~Re clustering occurs without any assistance from RED or RIP, and hence without the need for defect sinks. This is again a remarkable feature of these alloys, confirmed in several studies \cite{herschitz1985radiation,2017ACTAXu,Klimenkov2016}. As noted by Herschitz and Seidman, ``The coherent precipitates were not associated with either linear or planar defects or with any impurity atoms; i.e. a true homogeneous radiation-induced precipitation occurs in this alloy'', or by Klimenkov {\it et al}., who~point out that ``The formation of Re-rich particles with a round shape was detected in the single crystal material. These particles were formed independently of cavities". We leave out heterogeneous precipitation at voids, as the evidence in the literature is conflicting at this stage: discounted in some works \cite{herschitz1985radiation,2017ACTAXu} and observed in others \cite{Klimenkov2016}.

\section*{Acknowledgements}
C.~H., Y.~Z., and J.~M.'s work has been supported by the US Department of Energy's Office of Fusion Energy Sciences, grant DE-SC0012774:0001. Computer time allocations at UCLA's IDRE Hoffman2 supercomputer are also acknowledged.
L.~G.~and P.~E.~acknowledge support from the Swedish Research Council in the form of a Young Researcher grant, and the European Research Council via a Marie Curie Career Integration Grant. Computer time allocations by the Swedish National Infrastructure for Computing at NSC (Link\"oping) and C3SE (Gothenburg) are gratefully acknowledged.

\appendix
\section{Size dependence of physical time in kMC simulations}\label{appendix}

As explained in Section \ref{sec:disc}, the mechanism of formation of Re clusters requires the concerted action of both interstitials and vacancies. In order to be able to capture their formation during reasonable computational times, the temperature regime considered must be one where the mobility of both species is comparable ($1700\sim2000$ K in our case). Then, the rate of arrival of solute atoms to a previously-nucleated Re cluster can be approximated by:
\begin{equation}\label{app0}
	r_{s}= \frac{1}{t_{\rm FP}+t_{diff}}
\end{equation}
where $t_{\rm FP}$ and $t_{diff}$ are the average time in between successive Frenkel-pair insertions and a characteristic diffusion time required by a vacancy and an interstitial to recombine with one another. $r_{s}$ is measured in units of atoms per unit time. At the temperatures and dose rates considered here, $t_{\rm FP}\gg t_{diff}$, such that $r_s\approx{t_{\rm FP}}^{-1}$. Assuming then that for each Frenkel pair inserted a minimum of one solute atom is transported:
\begin{equation}\label{app1}
	r_{s}= \frac{dN_B}{dt}=r_{\rm dpa}N
\end{equation}
where $N_B$ is the total number of solute atoms in the precipitate. $r_{\rm dpa}$ in the above equation is the damage rate, expressed in units of [dpa$\cdot$s$^{-1}$]. The precipitate volume growth rate is directly equal to the atomic volume times $r_{s}$:
\begin{equation}
	\dot{V}_{\rm ppt}=\Omega_ar_{s}= \Omega_a\frac{dN_B}{dt}=\Omega_ar_{\rm dpa}N
\end{equation}
Assuming that the precipitate is close to spherical:
$$\dot{V}_{\rm ppt}=4\pi {R_{\rm ppt}}^2\dot{R}_{\rm ppt}=\Omega_ar_{\rm dpa}N$$
And, operating, we arrive at the equation for the evolution of the precipitate radius with time:
\begin{equation}
R_{\rm ppt}=\left( \frac{\Omega_a r_{\rm dpa}Nt}{4\pi} \right)^{\frac{1}{3}}
\end{equation}
which is the equation used for fitting in Fig.\ \ref{radius}.

Then, from eq.\ \eref{app1}, for a given constant dpa rate, it is clear that the ratio $r_s(V_1)N_1^{-1}=r_s(V_2)N_2^{-1}=~{\rm constant}$, where $V_1$ and $V_2$ are two different box sizes. For as long as the approximation in eq.\ \eref{app0} is valid, then:
$$t_{\rm FP}^{\rm (1)}N_1=t_{\rm FP}^{\rm (2)}N_2=~{\rm constant}$$
which allows us to compare simulations done on box sizes of $64^3$ and $80^3$ directly. We emphasize that at lower temperatures, and/or high dose rate, where $t_{\rm FP}\approx t_{diff}$, this comparison is no longer valid.

\section*{References}

\bibliography{ref}

\end{document}